\RequirePackage[hyphens]{url}
\documentclass[fleqn,usenatbib]{mnras}

\usepackage{newtxtext,newtxmath}

\usepackage{hyperref}
\usepackage{natbib}
\usepackage[dvipsnames]{xcolor}
\usepackage[utf8]{inputenc}
\usepackage{ctable}

\usepackage[T1]{fontenc}

\newcommand{\Mhalo}{M_{\rm{halo}}}
\newcommand{\Mcgm}{M_{\rm{cgm}}}
\newcommand{\fcgm}{f_{\rm{cgm}}}
\newcommand{\fcool}{f_{\rm{cool}}}
\newcommand{\logT}{\log(T_{\rm{cgm}})}
\newcommand{\logZ}{\log(Z_{\rm{cgm}})}
\newcommand{\hmpc}{h^{-1}{\rm Mpc}}

\DeclareRobustCommand{\VAN}[3]{#2}
\let\VANthebibliography\thebibliography
\def\thebibliography{\DeclareRobustCommand{\VAN}[3]{##3}\VANthebibliography}

\usepackage{graphicx}	
\usepackage{amsmath}	
\usepackage{array}
\usepackage[verbose]{placeins}
\usepackage{dutchcal}
\usepackage{multirow}

\title[CGM with CNN]{An Observationally Driven Multifield Approach for Probing the Circum-Galactic Medium with Convolutional Neural Networks}

\author[N. Gluck et al.]{\noindent
Naomi Gluck$^{1}$\thanks{E-mail: naomi.gluck@yale.edu}, Benjamin D. Oppenheimer$^{2}$, Daisuke Nagai$^{1}$, Francisco Villaescusa-Navarro$^{3,4}$, \newauthor Daniel Anglés-Alcázar$^{5,4}$ \\
$^{1}$Physics Department, Yale University, 217 Prospect St, New Haven, CT, 06511, USA\\
$^{2}$CASA, Department of Astrophysical and Planetary Sciences, University of Colorado, 389 UCB, Boulder, CO 80309, USA\\
$^{3}$Department of Astrophysical Sciences, Princeton University, Peyton Hall, Princeton, NJ, 08544, USA\\
$^{4}$Center for Computational Astrophysics, Flatiron Institute, 162 5th Avenue, New York, NY, 10010, USA\\
$^{5}$Department of Physics, University of Connecticut, 196 Auditorium Road, U-3046, Storrs, CT, 06269, USA
}

\date{Accepted 2023 December 5. Received 2023 December 3; in original form 2023 September 12}

\pubyear{2023}

\begin{document}
\label{firstpage}
\pagerange{\pageref{firstpage}--\pageref{lastpage}}
\maketitle

\begin{abstract}
The circum-galactic medium (CGM) can feasibly be mapped by multiwavelength surveys covering broad swaths of the sky. With multiple large datasets becoming available in the near future, we develop a likelihood-free Deep Learning technique using convolutional neural networks (CNNs) to infer broad-scale physical properties of a galaxy’s CGM and its halo mass for the first time. Using CAMELS (Cosmology and Astrophysics with MachinE Learning Simulations) data, including IllustrisTNG, SIMBA, and Astrid models, we train CNNs on Soft X-ray and 21-cm (HI) radio 2D maps to trace hot and cool gas, respectively, around galaxies, groups, and clusters. Our CNNs offer the unique ability to train and test on ``multifield'' datasets comprised of both HI and X-ray maps, providing complementary information about physical CGM properties and improved inferences. Applying eRASS:4 survey limits shows that X-ray is not powerful enough to infer individual halos with masses $\log(\Mhalo/M_{\odot}) < 12.5$. The multifield improves the inference for all halo masses. Generally, the CNN trained and tested on Astrid (SIMBA) can most (least) accurately infer CGM properties. Cross-simulation analysis -- training on one galaxy formation model and testing on another -- highlights the challenges of developing CNNs trained on a single model to marginalize over astrophysical uncertainties and perform robust inferences on real data. The next crucial step in improving the resulting inferences on the physical properties of CGM depends on our ability to interpret these deep-learning models.

\end{abstract}



\section{Introduction}
\label{sec:intro}

New telescopes are currently engaged in comprehensive surveys across large sky areas and reaching previously unobtainable depths, aiming to map the region beyond the galactic disk but within the galaxy's virial radius: the circum-galactic medium, or the CGM \citep{tumlinson_circumgalactic_2017}. However, these telescopes have inherent limitations in detecting emissions from gaseous halos surrounding typical galaxies. Nevertheless, they offer an exceptional opportunity to characterise the broad properties of CGM that extend beyond their original scientific scope. The CGM contains a multiphase gas, partly accreted from the filaments of the cosmic web that is continuously being reshaped, used in star formation, and enriched by astrophysical feedback processes occurring within the galaxy \citep{Keres_2005, oppenheimer_2016,christensen_2016,alcazar_fire_2017, hafen_2019}.

A simple way to characterise the CGM is by temperature.  The cool phase gas has a temperature of approximately $T\sim 10^4$ K and has been the focus of UV absorption line measurements \citep[e.g.,][]{cooksey_2010,tumlinson_2013,werk_2013,Johnson15,keeney_2018}.  The hot phase of the CGM, with temperatures $T > 10^6$ K, is observable via X-ray facilities \citep[e.g.,][]{bregman_2018,bogdan_groups_2018, mathur_2023} and can contain the majority of a galaxy's baryonic content. Understanding both the cool and hot phases of the CGM may answer questions regarding where we may find baryons \citep{anderson_2011,werk_2014,Li_spirals_2017,Oppenheimer_2018_multiphase_cgm}, how galaxy quenching proceeds \citep{tumlinson_2011,somerville_2015}, and how the metal products of stellar nucleosynthesis are distributed \citep{peebles_2014}.

New, increasingly large datasets that chart the CGM across multiple wavelengths already exist. In particular, two contrasting wavelengths map diffuse gas across nearby galaxies: the X-ray and the 21-cm (neutral hydrogen, HI) radio. First, the \textit{eROSITA}\footnote{Although {\it eROSITA} is currently dormant; its data at the level we mock have already been taken.} 
mission has conducted an all-sky X-ray survey, enabling the detection of diffuse emission from hot gas associated with groups and clusters and potentially massive galaxies \citep{predehl_erosita_2021}. Second, in the 21-cm radio domain, the pursuit of detecting cool gas encompasses initiatives that serve as precursors to the forthcoming Square Kilometer Array (SKA) project. Notable among these are ASKAP \citep{Johnston_2007} and MeerKAT \citep{jonas_meerkat_2016}, both of which have already conducted comprehensive surveys of HI gas in galaxy and group environments through deep 21-cm pointings.

Cosmological simulations provide theoretical predictions of CGM maps, yet divergences arise due to varying hydrodynamic solvers and subgrid physics modules employed in galaxy formation simulations \citep{somerville_2015, tumlinson_circumgalactic_2017,dave_galaxy_2020}. As a result, we see very different predictions for the circumgalactic reservoirs surrounding galaxies. Distinctively, the publicly available simulations such as IllustrisTNG \citep{nelson_first_2018, pillepich_simulating_2018}, SIMBA \citep{dave_simba_2019}, Astrid \citep{bird_astrid_2022, ni_astrid_2022}, among others \citep[e.g.,][]{schaye_eagle_2015,hopkins_fire2_2018,wetzel_fire_2023}, are valuable resources for generating CGM predictions. CAMELS\footnote{CAMELS Project Website: \url{https://www.camel-simulations.org}}\footnote{CAMELS Documentation available at \url{https://camels.readthedocs.io/en/latest/index.html}}  (Cosmology and Astrophysics with MachinE Learning Simulations) is the first publicly available suite of simulations that includes thousands of parameter and model variations designed to train machine learning models \citep{camels_2021_w_ML,camels_2021_fundamental_params}. It contains four different simulations \textit{sets} covering distinct cosmological and astrophysical parameter distributions: LH (Latin Hypercube, 1,000 simulations), 1P (1-Parameter variations, 61 simulations), CV  (Cosmic Variance, 27 simulations), and EX (Extreme, 4 simulations). Of these, the CV set is uniquely significant as it fixes cosmology and astrophysics to replicate the observable properties of galaxies best, providing a fiducial model. We exclude the numerous CAMELS simulations that vary cosmology and astrophysical feedback to prevent unrealistic galaxy statistics. Thus, utilising the diverse CAMELS CV sets, we explore three universe realisations that make distinguishing predictions for the CGM.

\begin{table}
\centering
\caption{Definitions and global value ranges of the CGM properties to be inferred and constrained by the network. These are the global value ranges, encompassing the individual ranges of IllustrisTNG, SIMBA, and Astrid. They remain consistent throughout any combination of simulations during training and testing. Properties are further distinguished by those radially defined by $R_{\rm 200c}$ and those by $200\ \rm{kpc}$.}
\label{table:new_params_and_ranges}
\begin{tabular}{p{0.055\textwidth}p{0.26\textwidth}>{\centering\arraybackslash}p{0.07\textwidth}}
 \specialrule{.1em}{.05em}{.05em} 
 \textbf{Property} & \textbf{Definition} & \textbf{Range} \\ [0.5ex] 
 \specialrule{.1em}{.05em}{.05em}
 $M_{\rm{halo}}$ & Logarithmic Halo mass in $R_{200c}$ & 11.5 -- 14.3  \\ 

 $f_{\rm{cgm}}$ & Mass ratio of CGM gas to total mass within $R_{200c}$ & 0.0 -- 0.23 \\
 \hline

 $Z_{\rm cgm}$ & Logarithmic CGM metallicity in 200 kpc & -3.6 -- -1.3  \\

 $M_{\rm{cgm}}$ & Logarithmic CGM mass in 200 kpc & 8.0 -- 12.5  \\

 $f_{\rm{cool}}$ & Ratio of cool, low-ionized CGM gas within 200 kpc & 0.0 -- 1.0   \\

 $T_{\rm cgm}$ & Logarithmic CGM temperature in 200 kpc & 3.9 -- 7.6  \\
 \specialrule{.1em}{.05em}{.05em}
\end{tabular}
\end{table}

In this study, we develop an image-based Convolutional Neural Network (CNN) to infer CGM properties from CAMELS IllustrisTNG, SIMBA, and Astrid CV-set simulations. The definitions and ranges for all CGM properties are outlined in Table~\ref{table:new_params_and_ranges}. Two significant and differently structured astrophysical feedback parameters that impact CGM properties, stellar and AGN feedback, remain predominantly unconstrained. The CV set does not explore the range of CAMELS feedback parameters like the other sets. However, we choose the CV set as a proof-of-concept and plan to include the much larger LH set that completely marginalizes over astrophysics \citep{villaescusa-navarro_robust_2021} in the future. The CNN is trained and tested on diverse simulations, yielding valuable insights into the CGM properties. Additionally, we apply observational multiwavelength survey limits to the CNN for each field, guiding the design and approach of new instruments and novel surveys, maximising their scientific returns on CGM properties, and significantly advancing our understanding of galaxy formation and large-scale structure.

This paper is outlined as follows.  \S\ref{sec:methods} lays out the methods used to complete this work and includes subsections on specific simulation information (\S\ref{sec:sim_info}), dataset generation (\S\ref{sec:dataset_gen}), CNNs (\S\ref{sec:cnn}), and network output (\S\ref{sec:network_outputs}). We begin \S\ref{sec:results} by presenting results using individual simulations to infer first the entire halo mass (\S\ref{sec:halo_mass}), then a global CGM property, the mass of the CGM over the mass of the halo, or $f_{\rm cgm}$ (\S\ref{sec:fcgm}), and the metallicity of the CGM (\S\ref{sec:logZ}) which exhibits large variation. We show results based on idealised soft X-ray and HI images and assess the impact of realistic observations with observational survey limits (\S\ref{sec:obslimit}). We also perform \textit{cross simulation inference}, where one trains a CNN on one galaxy formation model or simulation and tests on another to gauge its robustness (\S\ref{sec:cross_inference}). We discuss the interpretability of the cross-simulation inference analysis (\S\ref{sec:cross_sec_interpret}), the applicability and limitations of CNNs applied to CGM (\S\ref{sec:limitations}), compare the variance between true and inferred values for CGM properties using the idealised multifield maps (\S\ref{sec:compare_variance}), and a possible avenue for future work as an expansion of this analysis (\S\ref{sec:future_work}). Lastly, \S\ref{sec:conclusions} concludes.

\section{Methods}
\label{sec:methods}

In this section, we introduce the simulations (\S\ref{sec:sim_info}) followed by how our halo-centric ``map'' datasets are generated and a description of the global properties we train the network to infer (\S\ref{sec:dataset_gen}). Then, \S\ref{sec:cnn} describes the neural network applied to these datasets. Finally, we specify the network output, including statistical measures, to evaluate the performance of CNN (\S\ref{sec:network_outputs}). 

We define some vocabulary and common phrases within this work. \textit{Fields} refer to X-ray and 21-cm HI (hereafter HI), where using one field corresponds to either X-ray or HI; two fields, X-ray and HI, make up the multifield. With our CNN architecture, the number of fields is equivalent to the number of channels. \textit{Parameters} and \textit{hyperparameters} define the inner workings of the CNN, where the latter must be optimised. This should not be confused with parameters in the context of astrophysical feedback. \textit{Properties} describe the attributes of the CGM that are inferred by the network: $\Mhalo$, $\fcgm$, $\logZ$, $\Mcgm$, $\fcool$, and $\logT$. The \textit{parameter space} reflects the range of values for the CGM properties (between the $16^{\rm th}$ and $84^{\rm th}$ percentiles) that each simulation encapsulates.

\subsection{Simulations}
\label{sec:sim_info}

\begin{figure*}
    \centering
    \includegraphics[scale=0.38]{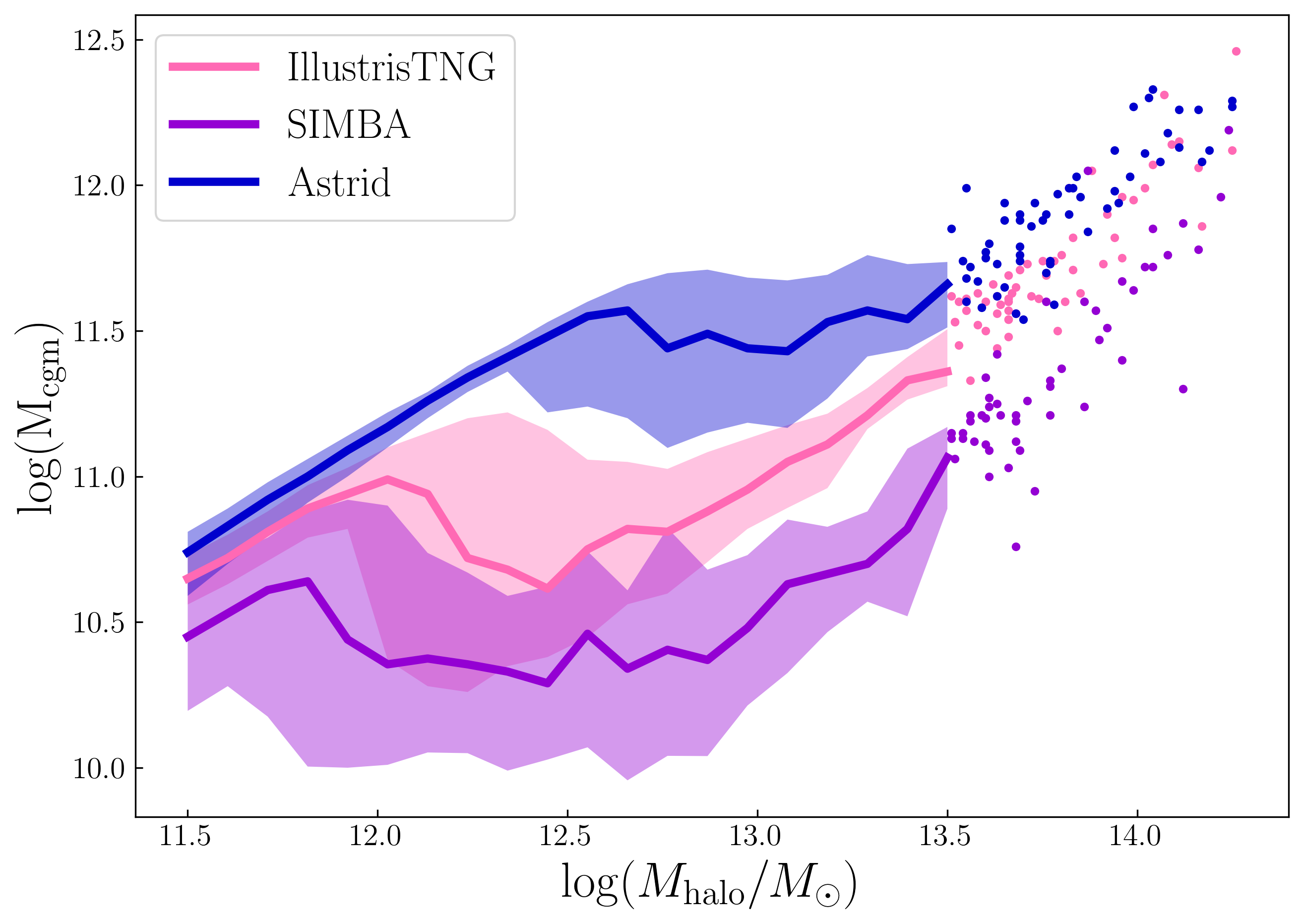}
    \includegraphics[scale=0.38]{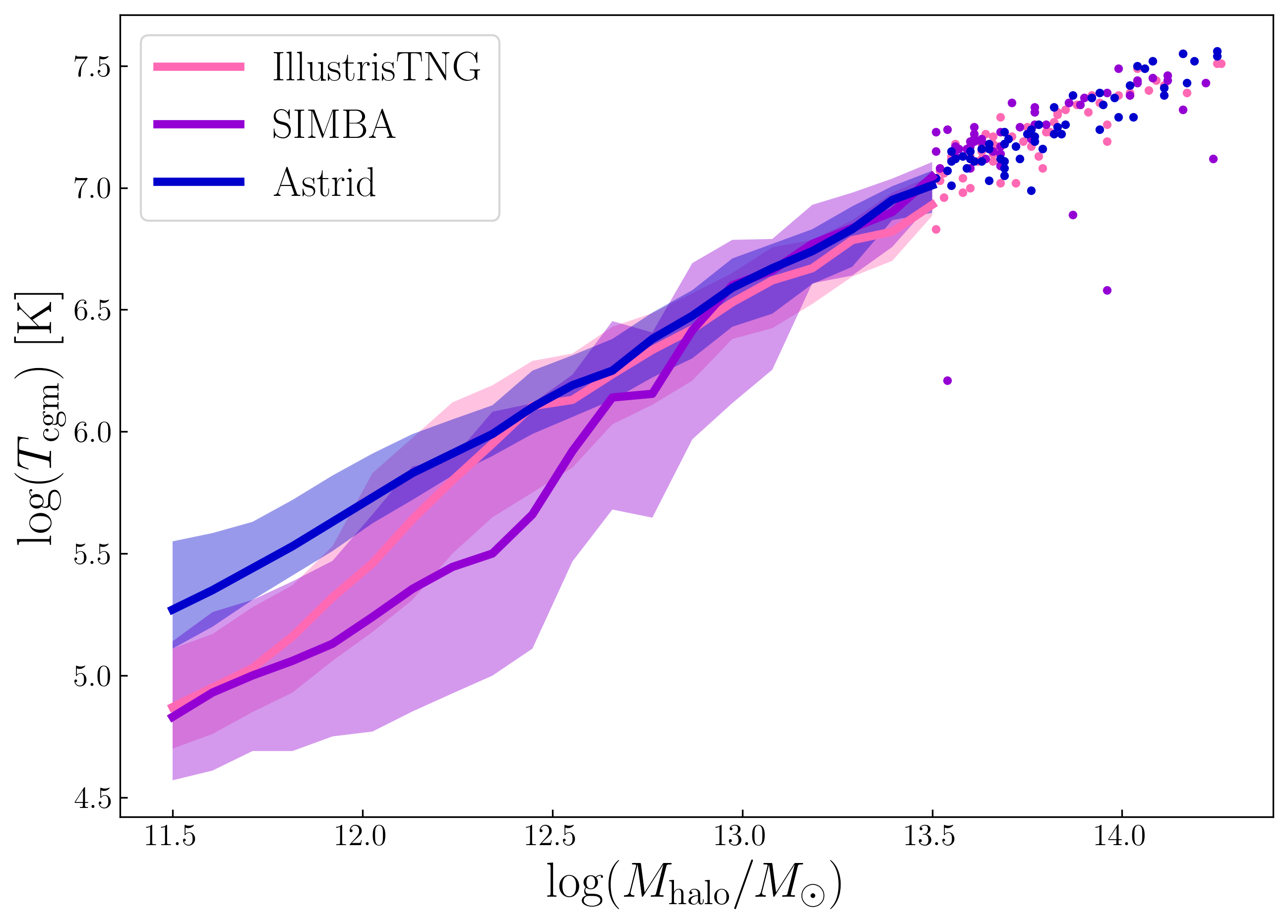}

    \includegraphics[scale=0.38]{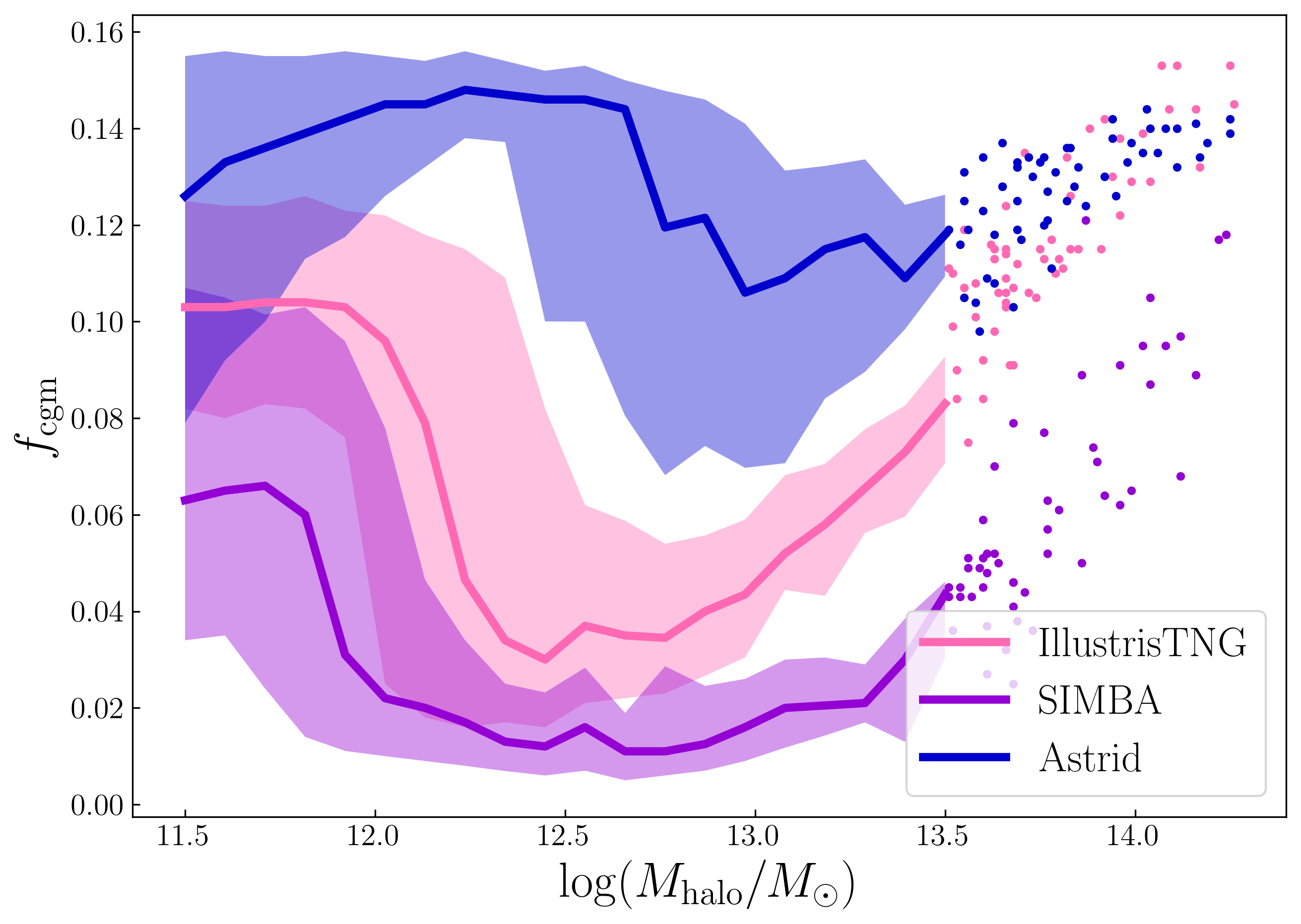}
    \includegraphics[scale=0.38]{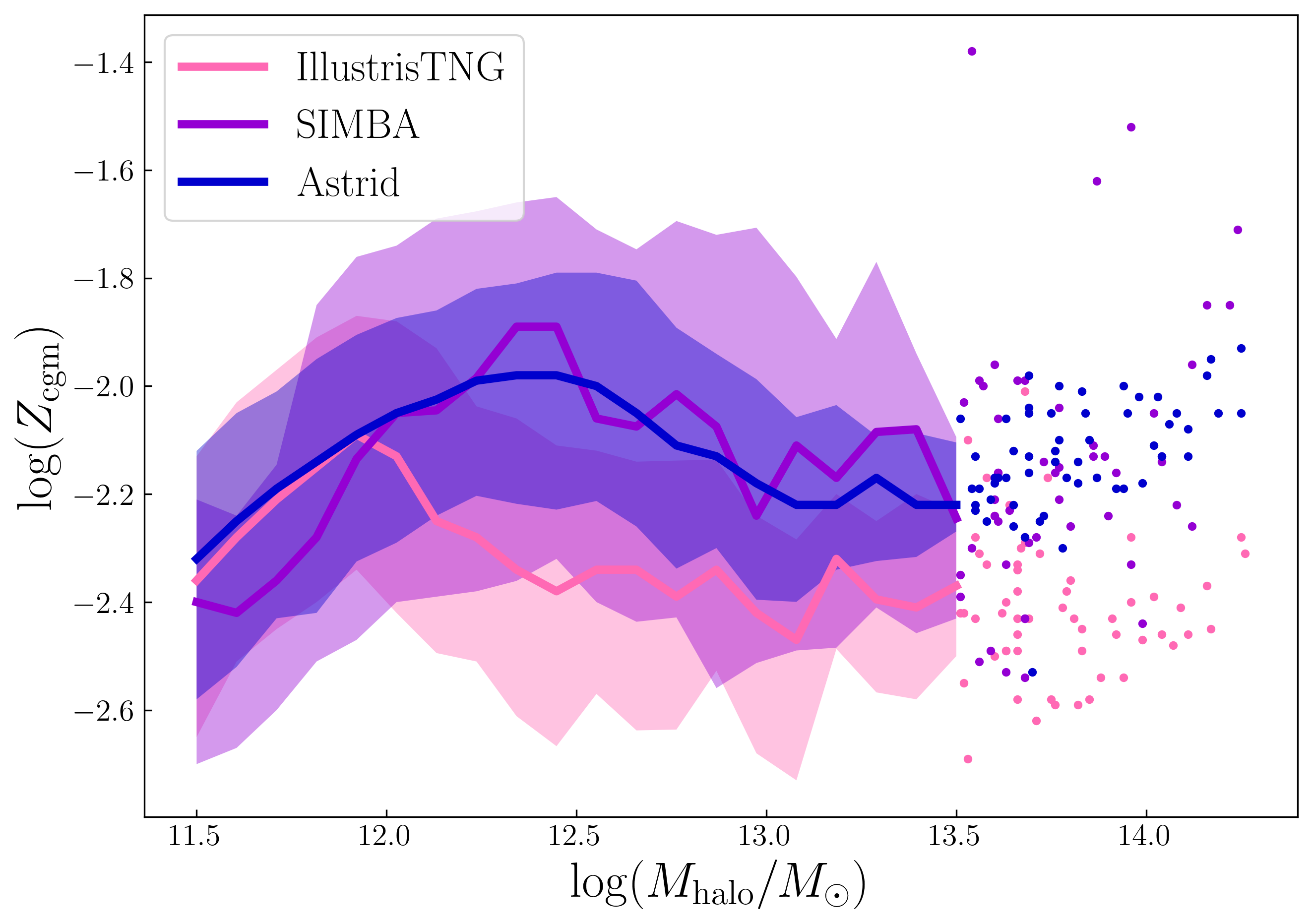}
    
    \caption{Relationship between different CGM properties and the halo mass. Panels specifically illustrate $\Mcgm$, $\logT$, $\fcgm$, and $\logZ$ within each simulation (IllustrisTNG, SIMBA, and Astrid) to represent the mean distribution of the objects. The points indicate the mass bins where there are statistically fewer halos in mass bins above $\log(\Mhalo/M_{\odot})=13.5$. The shaded regions represent the $16^{\rm th}-84^{\rm th}$ percentiles. }
    \label{fig:counts}
\end{figure*}

We use the CV (``Cosmic Variance'') set from three simulation suites, each of which uses a different hydrodynamic scheme: CAMELS-IllustrisTNG (referred to as IllustrisTNG) using AREPO \citep{springel_2010_arepo,weinberger_arepo_2020}, CAMELS-SIMBA (referred to as SIMBA) utilizing GIZMO \citep{hopkins_2015}, and CAMELS-Astrid (referred to as Astrid) utilizing MP-Gadget \citep{gadget_05}. These simulations encompass 27 volumes spanning $(25 \hmpc)^3$ with fixed cosmological parameters ($\Omega_{\rm M} = 0.3$ and $\sigma_{8}=0.8$) with varying random seeds for each volume's initial condition. The CAMELS astrophysical parameters for feedback are set to their fiducial values. We exclusively use the $z=0$ snapshots for this work. 

IllustrisTNG is an adaptation of the original simulation as described in \citet{nelson_illustristng_2019} and \citet{pillepich_simulating_2018}, using the AREPO \citep{springel_2010_arepo} magnetohydrodynamics code employing the N-body tree-particle-mesh approach for solving gravity and magnetohydrodynamics via moving-mesh methods. Like all simulation codes used here, IllustrisTNG has subgrid physics modules encompassing stellar processes (formation, evolution, and feedback) and black hole processes (seeding, accretion, and feedback). Black hole feedback uses a dual-mode approach that applies thermal feedback for high-Eddington accretion rates and kinetic feedback for low-Eddington rate accretion rates. The kinetic mode is directionally pulsed and is more mechanically efficient than the thermal mode \citep{weinberger_17}.

SIMBA, introduced in \cite{dave_simba_2019} uses the hydrodynamics-based ``Meshless Finite Mass'' GIZMO code \citep{hopkins_2015,hopkins_gizmo_2017}, with several unique subgrid prescriptions. It includes more physically motivated implementations of 1) AGN feedback and 2) black hole growth. SIMBA's improved subgrid physics model for AGN feedback is based on observations, utilising kinetic energy outflows for both radiative and jet feedback modes operating at high and low Eddington ratios, respectively. Additionally, it applies observationally motivated X-ray feedback to quench massive galaxies. SIMBA's black hole growth model is phase-dependent. Cool gas accretion onto BHs is achieved through a torque-limited accretion model \citep{angles_alcazar_2017a}, and when accreting hot gas, SIMBA transitions to Bondi accretion.

Astrid, introduced in \citet{bird_astrid_2022}, adopts the Pressure-Entropy SPH hydrodynamic model that uses the MP-Gadget code \citep{yu_feng_2018_1451799}. The original Astrid simulations focus on modelling high redshift galaxy formation (from z=99 to z=3) by considering inhomogeneous hydrogen and helium reionization, metal return from massive stars, and the initial velocity offset between baryons and dark matter. It has also enhanced the modelling of black hole mergers via a dynamic friction model. The CAMELS version of Astrid \citep{ni_camels_2023} follows the original simulation, but slight changes in black-hole dynamics and dual-mode AGN feedback implementations were made between them.

\begin{table}
\centering
\caption{Outlining the number of halos per mass bin in IllustrisTNG, SIMBA, and Astrid. The mass bins are defined as follows: Sub-L* for small halos with mass between $11.5 \le \log(\Mhalo/M_{\odot}) \le 12$, L* for intermediate-sized halos with masses ranging from $12 \le \log(\Mhalo/M_{\odot}) \le 13$, and Groups are large halos with masses from $13 \le \log(\Mhalo/M_{\odot})\le 14.3$.}
\label{table:mass_bins}
\begin{tabular}{l c c c c}
 \specialrule{.1em}{.05em}{.05em}
 \textbf{Simulation} & \textbf{Sub-L*} & \textbf{L*} & \textbf{Group} & \textbf{Total} \\ [0.5ex] 
 \specialrule{.1em}{.05em}{.05em}
 IllustrisTNG & 3450 & 1812 & 192 & 5454  \\ 
 SIMBA & 3397 & 1534 & 170 & 5101 \\
 Astrid & 3262 & 1866 & 218 & 5346 \\ 
 \specialrule{.1em}{.05em}{.05em}
\end{tabular}
\end{table}

\subsection{Dataset Generation}
\label{sec:dataset_gen}

To create our halo-centric map datasets, we use {\tt yt}-based software \citep{turk_yt_2011} that allows for consistent and uniform analysis across different simulation codes. We generate maps of all halos within the CV set with masses of at least $M_{\rm halo}=10^{11.5}\;{\rm M}_{\odot}$ along the three cardinal axes.  There are approximately 5,000 halos for each simulation. The highest halo mass is $10^{14.3}\;{\rm M}_{\odot}$, for a nearly 3~dex span in halo mass. Refer to Table~\ref{table:mass_bins} for additional details. We categorise all the halos within the simulations by halo mass, where Sub-L* halos are within the range $11.5 \le \log(M_{\rm{halo}}/M_{\odot}) \le 12$, L* halos are within the range $12 \le \log(M_{\rm{halo}}/M_{\odot}) \le 13$, and groups are within the range $ 13 \le \log(M_{\rm{halo}}/M_{\odot}) \le 14.3$.

The relationship between $\log(\Mhalo/M_{\odot})$ and $\log(\Mcgm)$, $\logT$, $\fcgm$, and $\logZ$ for all simulations, the parameter space, is shown in Figure~\ref{fig:counts}. The mean value of each property is indicated with a solid line. The shaded regions represent the $16^{\rm{th}} - 84^{\rm th}$ percentiles, and the dotted points indicate the ``statistically low'' region for halos with halo masses above $\log(\Mhalo/M_{\odot})>13.0$. In agreement with previous work \citep{oppenheimer_simulating_2021,delgado_2023,ni_camels_2023,gebhardt_2023}, we illustrate how the properties of gas beyond the galactic disk can differ significantly between feedback implementations. 

For $\log(\Mcgm)$ (top left), Astrid (blue) shows little scatter below $\log(\Mhalo/M_{\odot}) > 12.5$, IllustrisTNG (pink) shows similar but less extreme scatter, and SIMBA (purple) has consistent scatter throughout. In $\logT$ (top right), Astrid again has a low scatter throughout the entire $\Mhalo$ range. This scatter increases slightly for IllustrisTNG and again for SIMBA, and it is interesting to note the divergence from $\logT \propto \log(\Mhalo/M_{\odot})^{2/3} $. Astrid has the most scatter for $\fcgm$ (bottom left), whereas IllustrisTNG and SIMBA display comparable scatter for lower masses, reducing for higher masses. Finally, $\logZ$ illustrates that all three simulations have significant and similar scatter. For $\Mcgm$, $\logT$, and $\fcgm$, Astrid has higher values throughout the $\Mhalo$ range, followed by IllustrisTNG and SIMBA. This is not the case in $\logZ$, where there is a significant overlap. The scatter in $\Mhalo$ was also computed with respect to the total flux per map, corresponding to the sum of all pixel values in X-ray and HI separately. When binned by $\Mhalo$, there are correlations only with IllustrisTNG and Astrid for X-ray (see Fig.~\ref{fig:mhalo_pix}). A more detailed discussion of map trends and pixel counts is in Appendix~\ref{sec:appendix_mock_datasets}.

From the snapshot data obtained from the X-ray and HI maps, we provide an equation describing the calculation of each CGM property ($\Mhalo$, $\fcgm$, $Z_{\rm cgm}$, $\Mcgm$, $\fcool$, and $T_{\rm cgm}$):  

\begin{gather}
    \begin{aligned}
            \Mhalo = &\sum m_{\rm DM}(r<R_{\rm 200c}) \\ &+\sum m_{\rm gas}(r<R_{\rm 200c})+\sum m_{\rm star}(r<R_{\rm 200c})
    \end{aligned}\label{eqn:mhalo}
\end{gather}

\begin{equation}
        \fcgm = \frac{\sum m_{{\rm cgm}}(r<R_{\rm 200c})}{\Mhalo}
        \label{eqn:fcgm}
\end{equation}

\begin{equation}
        Z_{\rm{cgm}} = \frac{\sum \mathcal{z}_{\rm{cgm}} (r<200\ \rm{kpc})}{\sum m_{\rm{cgm}} (r<200\ \rm{kpc})}
        \label{eqn:Zcgm}
\end{equation}

\begin{equation}
        \Mcgm = \sum m_{\rm cgm}(r<200 {\rm kpc})
        \label{eqn:Mcgm}
\end{equation}

\begin{equation}
        \fcool = \frac{\sum m_{\rm cool}(r < 200\ \rm{kpc})}{\sum m_{\rm cgm}(r < 200\ \rm{kpc})}
        \label{eqn:fcool}
\end{equation}

\begin{equation}
        T_{\rm{cgm}} = \frac{\sum \mathcal{t}_{\rm{cgm}}(r<200\ \rm{kpc})}{ \sum m_{\rm{cgm}}(r<200\ \rm{kpc})}
        \label{eqn:Tcgm}
\end{equation}
where $m$ is the mass of dark matter (DM), gas, or stellar (star) particles enclosed within $r<200\ \rm{kpc}$. 
The subscript ``cgm'' refers to any gas that is not star-forming. $\mathcal{z}_{\rm cgm}$ is the metallicity of the gas particle. $m_{\rm cool}$ is CGM gas with $T<10^{6}$ K. $\mathcal{t}_{\rm cgm}$ is the temperature of the gas particle. For the definitions and numerical ranges of the above CGM properties, see Table~\ref{table:new_params_and_ranges}. To ensure our CNN is able to reproduce the scatter seen in Fig.~\ref{fig:counts}, we include a comparison of mean and variance values between the input parameters and the output inference, separated by mass bin for each galaxy simulation model. This is only computed for $\logZ$, as this parameter does not have a clear relationship with halo mass, enabling a distinction between mass relationships (seen within the other properties) and intrinsic scatter. We confirm that our CNN reproduces the scatter within the initial datasets.

We generated one channel for each field (HI or X-ray), adding them together in the multifield case (HI+X-ray). Each map utilises values obtained through mock observation, as described below. For X-ray, we map X-ray surface brightness emission in the soft band between 0.5 and 2.0 keV. HI, or ``Radio'' refers to the 21-cm emission-based measurement that returns column density maps, which is a data reduction output of 21-cm mapping techniques. Each map is 128x128 pixels, spanning $512\times512$ kpc$^2$ with a 4~kpc resolution. The depth spans $\pm 1000$~kpc from the centre of the halo. Two types of maps are generated for each field: those with no observational limits, called idealised maps, and those with observed limits imposed. We first explain the generation of idealised maps. X-ray maps are created using the pyXSIM package \citep{zuhone_pyxsim_2016}. While pyXSIM can generate lists of individual photons, we use it in a non-stochastic manner to map the X-ray emission across the kernel of the fluid element. Therefore, our X-ray maps are idealised in their ability to map arbitrarily low emission levels. Radio-based HI column density maps are created using the Trident package \citep{hummels_trident_2017} where the \citet{haardt12} ionisation background is assumed with the self-shielding criterion of \citet{rahmati13} applied.

\begin{figure}

\includegraphics[width=0.158\textwidth]{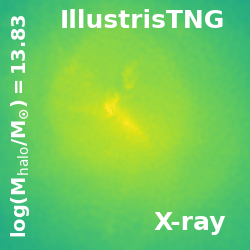}
\includegraphics[width=0.158\textwidth]{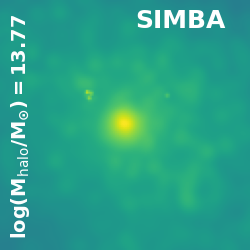}
\includegraphics[width=0.158\textwidth]{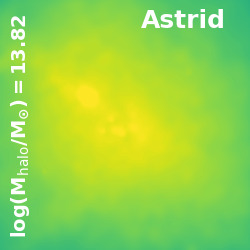}

\includegraphics[width=0.158\textwidth]{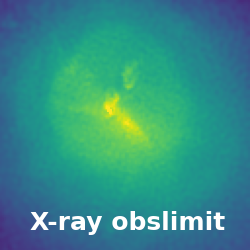}
\includegraphics[width=0.158\textwidth]{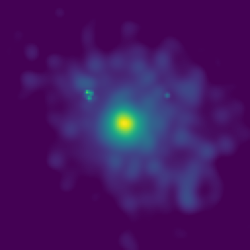}
\includegraphics[width=0.158\textwidth]{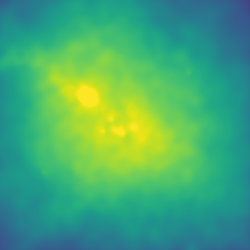}

\includegraphics[width=0.158\textwidth]{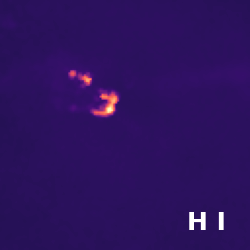}
\includegraphics[width=0.158\textwidth]{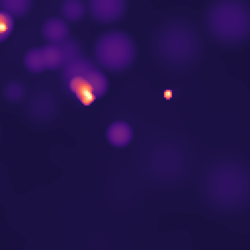}
\includegraphics[width=0.158\textwidth]{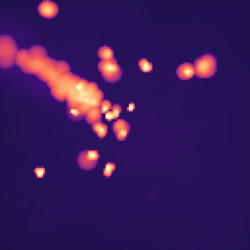}

\includegraphics[width=0.158\textwidth]{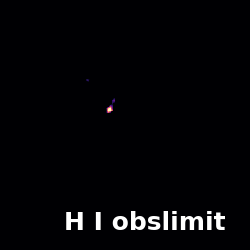}
\includegraphics[width=0.158\textwidth]{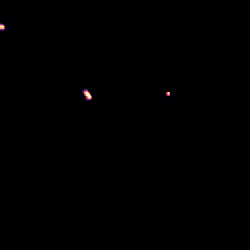}
\includegraphics[width=0.158\textwidth]{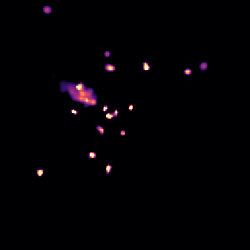}

\caption{Each column illustrates maps for IllustrisTNG, SIMBA, and Astrid, respectively. Each row corresponds to maps for idealised X-rays, X-rays with observational limits, idealized HI, and HI with observational limits. These maps display the same halo across the CV set of the three simulations.}
\label{fig:all_sim_maps}
\end{figure}

Figure~\ref{fig:all_sim_maps} depicts maps of the same massive halo in the three simulations: IllustrisTNG, SIMBA, and Astrid, from left to right, respectively. The four rows illustrate 1) idealised X-ray, 2) observationally limited X-ray, 3) idealized HI, and 4) observationally limited HI. For X-ray with observational limits, we set the surface brightness limit to $2.0\times 10^{-13}$ erg s$^{-1}$ cm$^{-2}$ arcmin$^{-2}$ corresponding to the observing depth of the {\it eROSITA} eRASS:4 all-sky survey \citep{predehl_erosita_2021}. For HI with observational limits, we set the column density limit to $N_{\rm HI}=10^{19.0}$ cm$^{-2}$, which is approximately the limit expected for the 21-cm HI MHONGOOSE Survey at a 15'' beam size similar to the {\it eROSITA} survey \citep{MeerKat_2016}. The observational limits are implemented by setting a lower limit floor that corresponds to the detectability of the telescope. Accessing the same halo across the three simulations is possible, since the CV set shares the same initial conditions between the different simulation suites. The X-ray maps tracing the gas primarily above $T>10^6$ K are brightest for Astrid and dimmest for SIMBA, a trend also seen when the observational limits are imposed. The HI maps, probing $T\sim 10^4$ K gas, are less centrally concentrated than X-ray and often trace gas associated with satellites.

We expand on the first column in Fig.~\ref{fig:all_sim_maps} in Fig.~\ref{fig:maps_fn_mass_v2}, formatted similarly, for a range of halo masses within IllustrisTNG from $\log(\Mhalo/M_{\odot})=13.83$ (leftmost) to $\log(\Mhalo/M_{\odot})=11.68$ (rightmost). X-ray emission, which traces the gas with a temperature above $10^6$ K, indicates a strong correlation with the halo mass. The features seen here include wind-blown bubbles \citep{predhel_2020_bubbles}, satellite galaxies that create bow shocks \citep{kang_2007_shocks,bykov_2008_shocks,zinger_2018_shocks,Li_2022_shocks}, and emissions associated with the galaxies themselves. HI does not have the same correlation with halo mass, strengthening our choice in creating the HI+X-ray multifield.

\subsection{Convolutional Neural Network}
\label{sec:cnn}

The advantage of employing CNNs lies in their capacity to simultaneously learn multiple features from various channels or fields (X-ray and HI). Fields can be used independently or together for training, validation, and testing without modifications to the network architecture and only minor changes in the scripts whenever necessary. This work adopts likelihood-free inference methods, suitable for cases where determining a likelihood function for large and complex datasets is computationally demanding or is not attainable. Our CNN architecture is based on the architecture used with the CAMELS Multifield Dataset (CMD) in \citet{villaescusanavarro2021multifield}, inferring two cosmological and four astrophysical feedback parameters. In addition to now inferring six CGM \textit{properties}, the remaining modifications stem from replacing the LH set \citep{camels_2021_fundamental_params} with the CV set. We must be able to accommodate unevenly distributed discrete-valued data in the form of ``halo-centric'' points, which can be seen in a property like the halo mass. This is compared to the LH-based CMD dataset, which was used to infer the evenly distributed cosmological and astrophysical feedback parameters by design.

Our CNN makes two main adjustments: 1) the kernel size is changed from 4 to 2 to accommodate a smaller initial network input, and 2) the padding mode is altered from ``circular'' to ``zeros.'' The padding mode is crucial in guiding the network when the image dimensions decrease, as it no longer perfectly fits the original frame. Changing to ``zeros'' means filling the reduced areas with zeros to maintain the network's functionality. The CNN architecture is outlined in greater detail in Appendix~\ref{sec:appendix_cnn_arch} in Table~\ref{table:architecture} for the main body of the CNN and Table~\ref{table:arch_end} for additional functions utilised after the main body layers.

 CNN also includes hyperparameters: 1) the \textit{maximum learning rate} (also referred to as step size), which defines how the application of weights changes during training, 2) the \textit{weight decay} as a regularisation tool to prevent overfitting by reducing the model complexity, 3) the \textit{dropout value} (for fully connected layers) as random neurone disablement to prevent overfitting, and 4) the \textit{number of channels} in CNN (set to an integer larger than one). To optimise these hyperparameters, we employ Optuna \citep{optuna_2019}\footnote{\url{https://github.com/pfnet/optuna/}}, a tool that efficiently explores the parameter space and identifies the values attributed to returning the lowest validation loss, thus achieving the best performance.

We divide the full dataset into a training set (60\%), a validation set (20\%) and a testing set (20\%). Only the training set contains the same halo along three different axis projections (setting the network parameter \texttt{split=3}). In contrast, the latter sets include neither the axis projections of any halo nor the original image of the halos assigned to the training set. The split is performed during each new training instance for a new combination of fields and simulations. We set the same random seed across all network operations, so as to drastically reduce the probability of overlap between training, validation, and testing sets. Without the same random seed, the dataset will not be split in the same way each time, and one halo could appear in two or more sets, causing inaccurate results. This process is necessary to ensure that the network does not perform additional ``learning'' in one phase.

\subsection{Network Outputs}
\label{sec:network_outputs}

Here, the ``moment'' network \citep{jeffrey2020} takes advantage of only outputting the mean, $\mu$, and variance, $\sigma$, of each property for increased efficiency, instead of a full posterior range. The minimum and maximum values used to calculate the network error for the six CGM properties are kept the same throughout this work, regardless of which simulation is used for training. Doing so ensures that the results are comparable in the cross-simulation analysis or training on one simulation and testing on another. 

We additionally include four metrics to determine the accuracy and precision of the CNN's outputs for each CGM property: the root mean squared error (RMSE), the coefficient of determination ($R^2$), the mean relative error ($\epsilon$), and the reduced chi-squared ($\chi^2$). In the formulae below, we use the subscript $i$ to correspond to the index value of the properties [1-6], the marginal posterior mean, $\mu_i$, and the standard deviation, $\sigma_i$. Four different statistical measurements are used to make such conclusions, and $\rm{TRUE}_i$ is used to denote the true value of any given CGM property with respect to simulation-based maps. 

\smallskip \noindent 
\textbf{Root mean squared error, RMSE}:
\begin{equation}
        \rm{RMSE}_i = \sqrt{\langle (\rm{TRUE}_i - \mu_i)^2\rangle}
        \label{eqn:rmse}
\end{equation}
where smaller RMSE values can be interpreted as increased model accuracy in units that can be related to the measured property. 

\smallskip \noindent 
\textbf{Coefficient of determination, $R^2$}:
\begin{equation}
        R_i^2 = 1 - \frac{\Sigma_i (\rm{TRUE}_i - \mu_i)^2}{\Sigma_i (\rm{TRUE}_i - \overline{\rm{TRUE}}_i)^2}
        \label{eqn:r_squared}
\end{equation}
representing the scale-free version of the RMSE, where the closer $R^2$ is to one, the more accurate the model. 

\smallskip \noindent 
\textbf{Mean relative error, $\epsilon$}:
\begin{equation}
        \epsilon_i = \left< \frac{\sigma_i}{\mu_i} \right>
        \label{eqn:epsilon}
\end{equation}
where smaller $\epsilon_i$ values can be interpreted as increased model precision. This is also the type of error predicted by CNN.

\smallskip \noindent 
\textbf{Reduced chi-squared, $\chi^2$}:
\begin{equation}
        \chi_i^2 = \frac{1}{N} \sum_{i=1}^{N} \left(\frac{\rm{TRUE}_i - \mu_i}{\sigma_i} \right)^2
        \label{eqn:chi2}
\end{equation}
where this quantifies how ``trustworthy'' the posterior standard deviation is, such that values close to one indicate a properly quantified error and that the model is well-trained. Values greater than one indicate that the errors are underestimated, and those smaller than one are overestimated. We do not expect deviations far from one when analysing inferences from CNNs trained and tested on the same simulation. However, values much higher than one are expected if network training and testing occur in different simulations, as outliers may have a large contribution. The variation in parameter spaces between simulations can be seen in Fig.~\ref{fig:counts}. If $\chi^2$ values become very large, two hypotheses can be presented. First, either the CNN is not powerful enough to output the correct inference from the provided maps, or second, there is not enough of the correct information within the dataset to produce a good inference. Distinguishing one hypothesis from the other, along with a physical interpretation of values that deviate from one, is not possible without the ability to interpret deep learning models. Resolving these issues is the focus of our future endeavours.

\begin{figure*}
    \centering
    \includegraphics[scale=0.6]{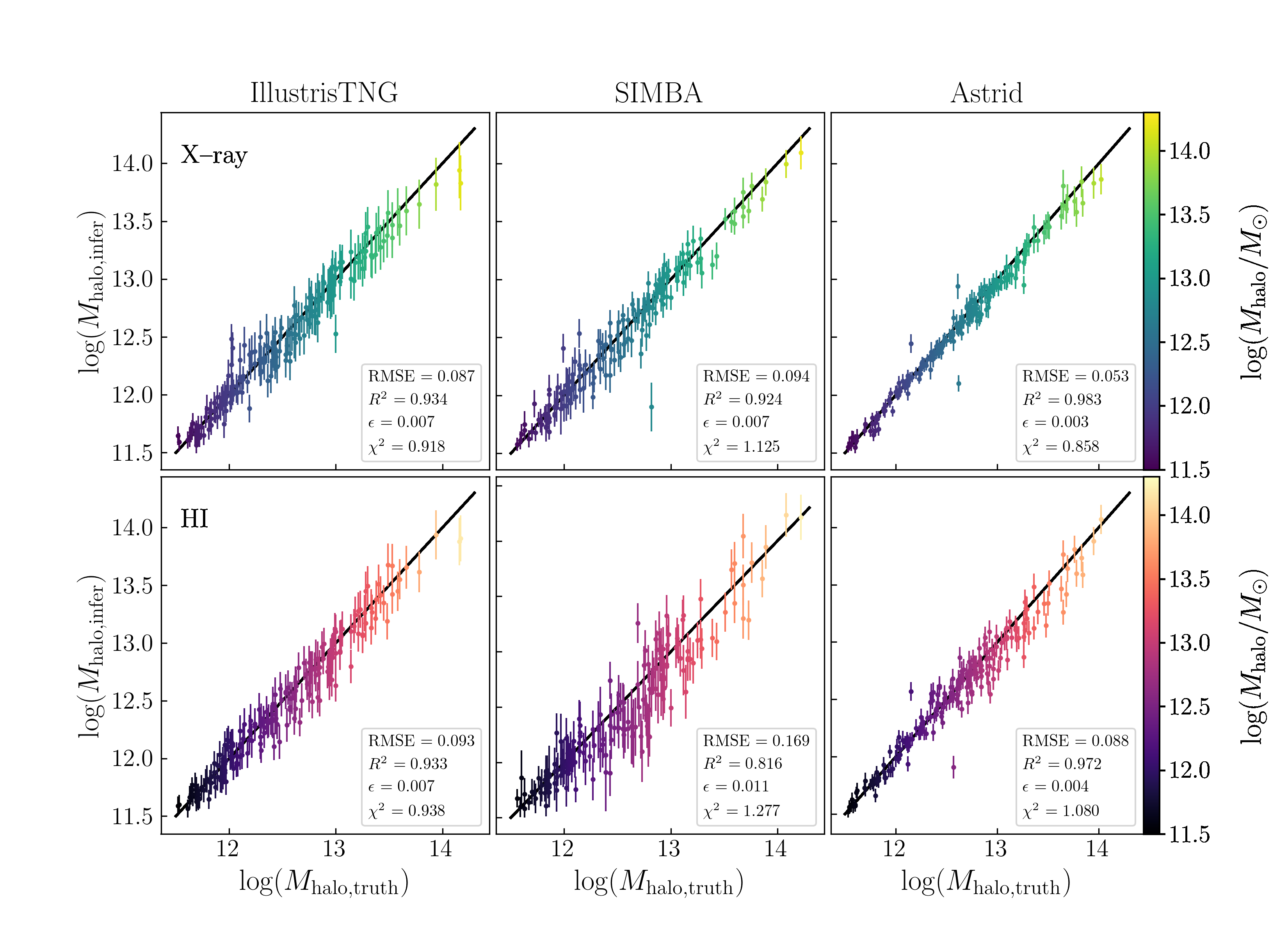}
    \caption{The Truth--Inference plots for $\Mhalo$ when training and testing on the same simulation using idealized single-field data. IllustrisTNG, SIMBA, and Astrid are shown from left to right, and X-ray and HI are shown in the upper and lower rows, respectively. The data are at $z=0.0$. We plot a mass-dependent fractional sample of halos from the testing set.}  
    \label{fig:Mhalo_idealized_fields}
\end{figure*}

It is also important to note that the values reported in the subsequent figures correspond to the subset of the data that has been plotted, not the entire set, unless otherwise noted. To achieve such a reduced set, we randomly select a fraction of data points that vary with halo mass -- for example, approximately $(1/30)^{\rm{th}}$ of $\log(\Mhalo/M_{\odot})=11.5$ halos, but all halos above $\log(\Mhalo/M_{\odot})= 13.0$ are plotted.

\section{Results}
\label{sec:results}

\begin{figure*}
    \centering
    \includegraphics[scale=0.6]{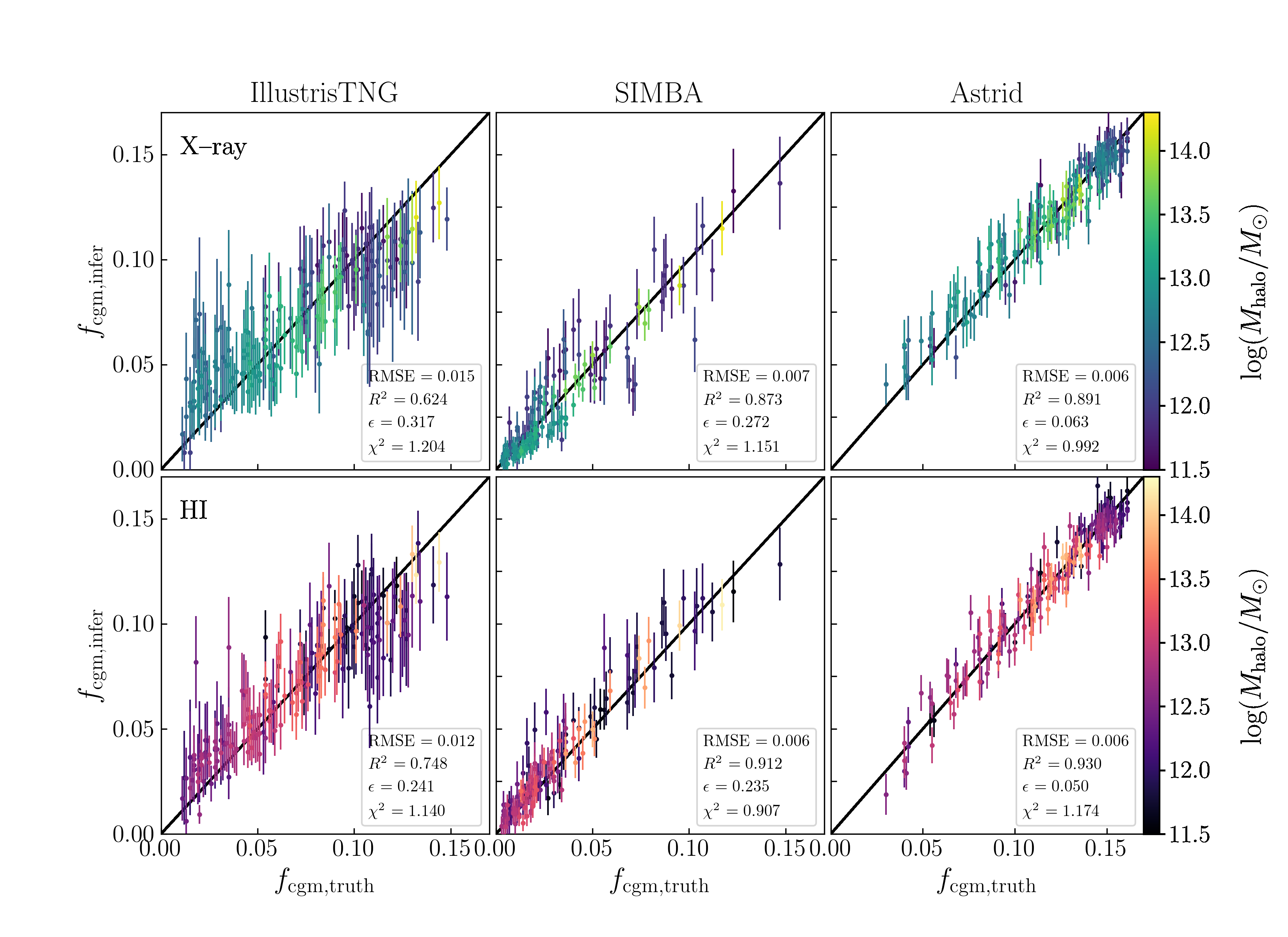}
    \caption{The Truth--Inference plots for $\fcgm$, with idealized X-ray (top) and idealized HI (bottom), where the color bar still represents halo mass. Astrid performs the best with the tightest constraints and the smallest errors, while IllustrisTNG performs the worst, likely due to the sharp increase of $f_{\rm{cgm}}$ at low mass. }
    \label{fig:fcgm_idealized_fields}
\end{figure*}

\begin{figure*}
    \centering
    \includegraphics[scale=0.6]{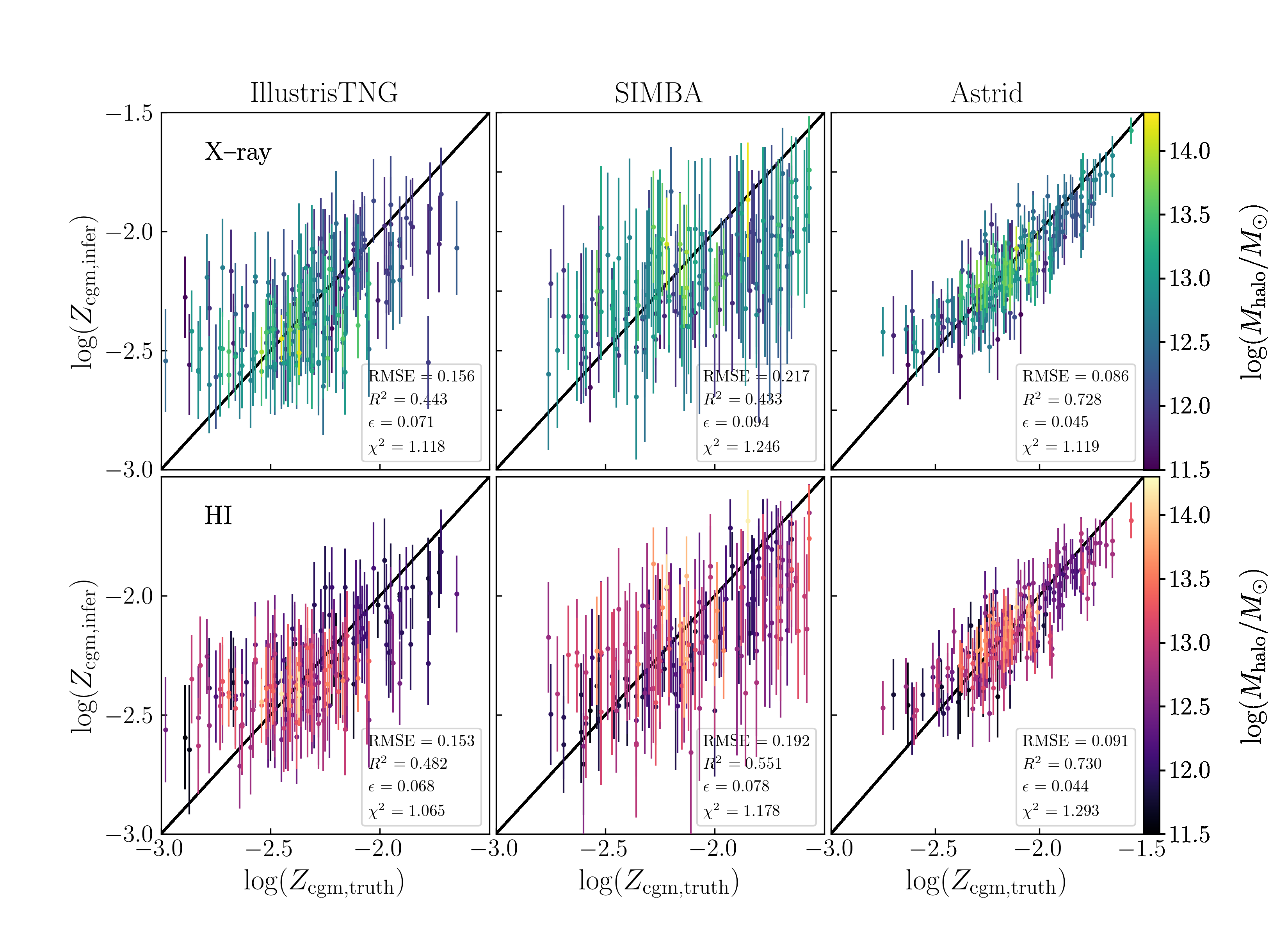}
    \caption{The Truth--Inference plots for metallicity, with idealised X-ray (top) and idealized HI (bottom), where the color bar still represents halo mass. Astrid performs the best, while SIMBA performs the worst, as it has the most varied Z values across the mass range, while Astrid has the most confined values.}
    \label{fig:logZ_idealized_fields}
\end{figure*}

\begin{figure*}
    \centering
    \includegraphics[scale=0.57]{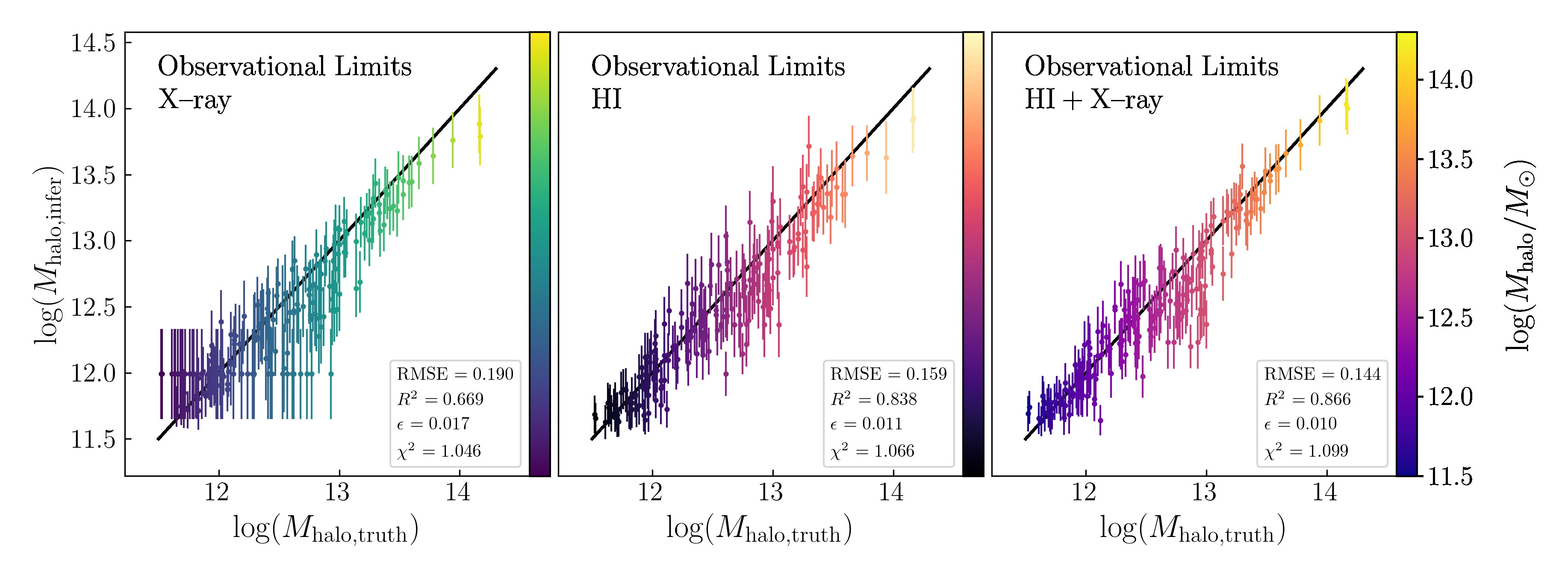}
    \includegraphics[scale=0.57]{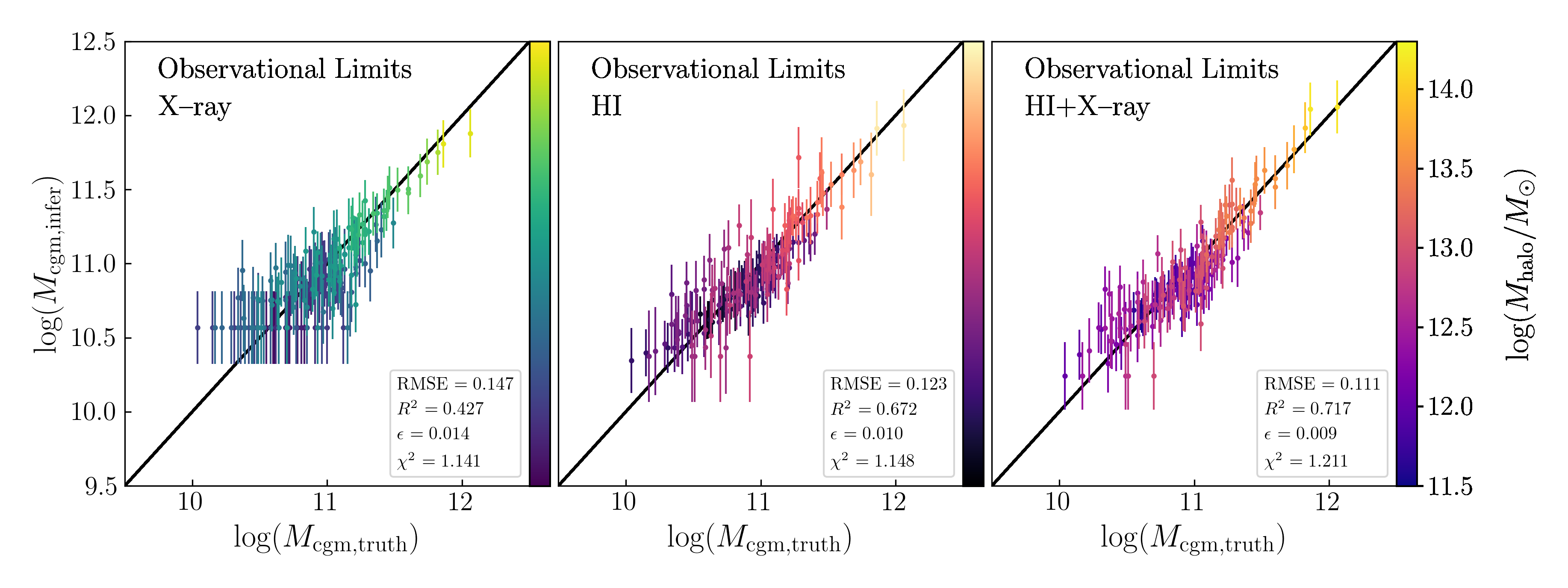}
    \caption{Truth--Inference figures for X-ray (left), HI (middle), and HI+X-ray (right) for IllustrisTNG with observational limits imposed on $\Mhalo$ (\textit{top row}) and $\Mcgm$ (\textit{bottom row}) using IllustrisTNG. X-ray provides poor inference, especially for lower-mass galaxies, as there are very few, sometimes no emission lines detected if they are too faint. On the other hand, the inference produced from HI results in more uniform errors throughout the mass range, since HI is detected around both low- and high-mass halos. Combined with their observational limits, the inference is enhanced by tighter constraints at all mass scales. }
    \label{fig:TNG_all_obslimits}
\end{figure*}

\begin{figure*}
    \centering
    \includegraphics[width=0.48\textwidth]{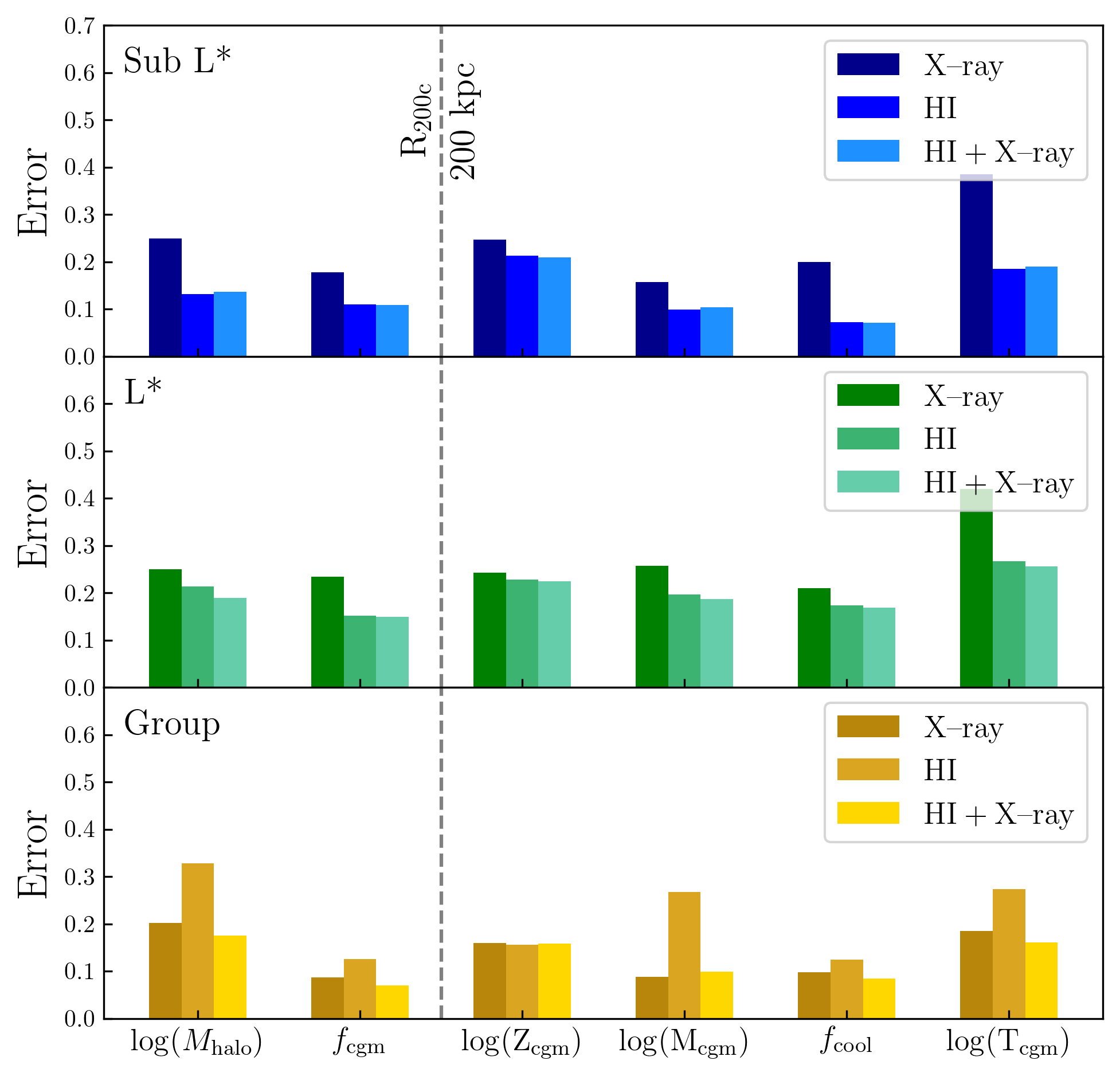}
    \includegraphics[width=0.48\textwidth]{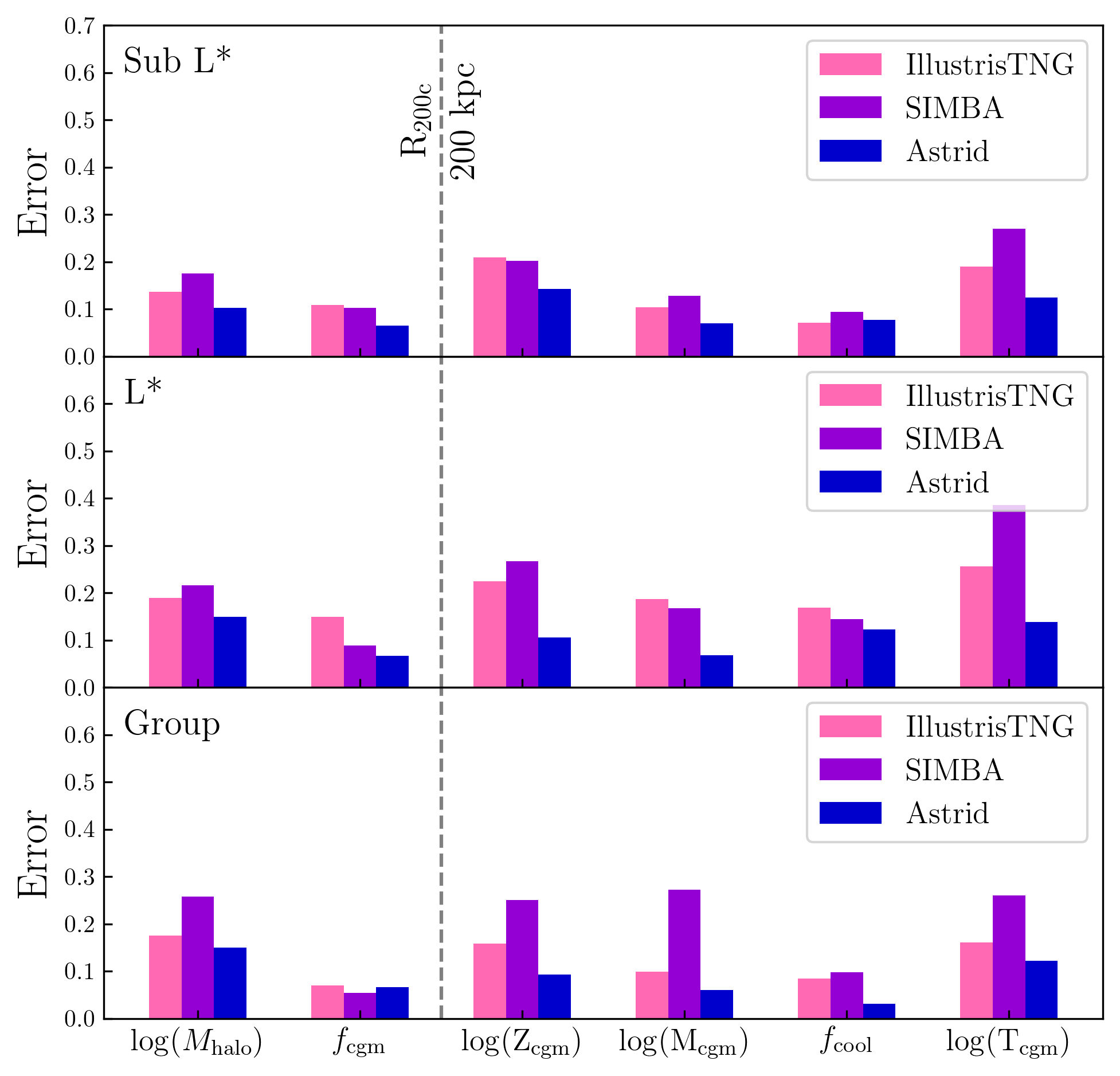}
    \caption{\textit{Left:} Average RMSE values split by halo category for training and testing on IllustrisTNG, with fields X-ray, HI, and the multifield HI+X-ray with observational limits, for all six properties. These bars are representative of the \textit{full} dataset. We provide a dashed vertical line to distinguish between properties that are radially bound by $R_{\rm 200c}$ and those by $200\ \rm{kpc}$. \textit{Right:} Average RMSE values split by simulation (training and testing on IllustrisTNG, SIMBA, or Astrid), with HI+X-ray and observational limits, for all six properties. These bars are representative of the \textit{full} dataset. Neither panel is entirely comparable to the Truth--Inference plots, as these categorise errors by halo mass and are for the full dataset.}
    \label{fig:err_obslimit_type}
\end{figure*}

\begin{figure*}
    \centering
    \includegraphics[scale=0.7]{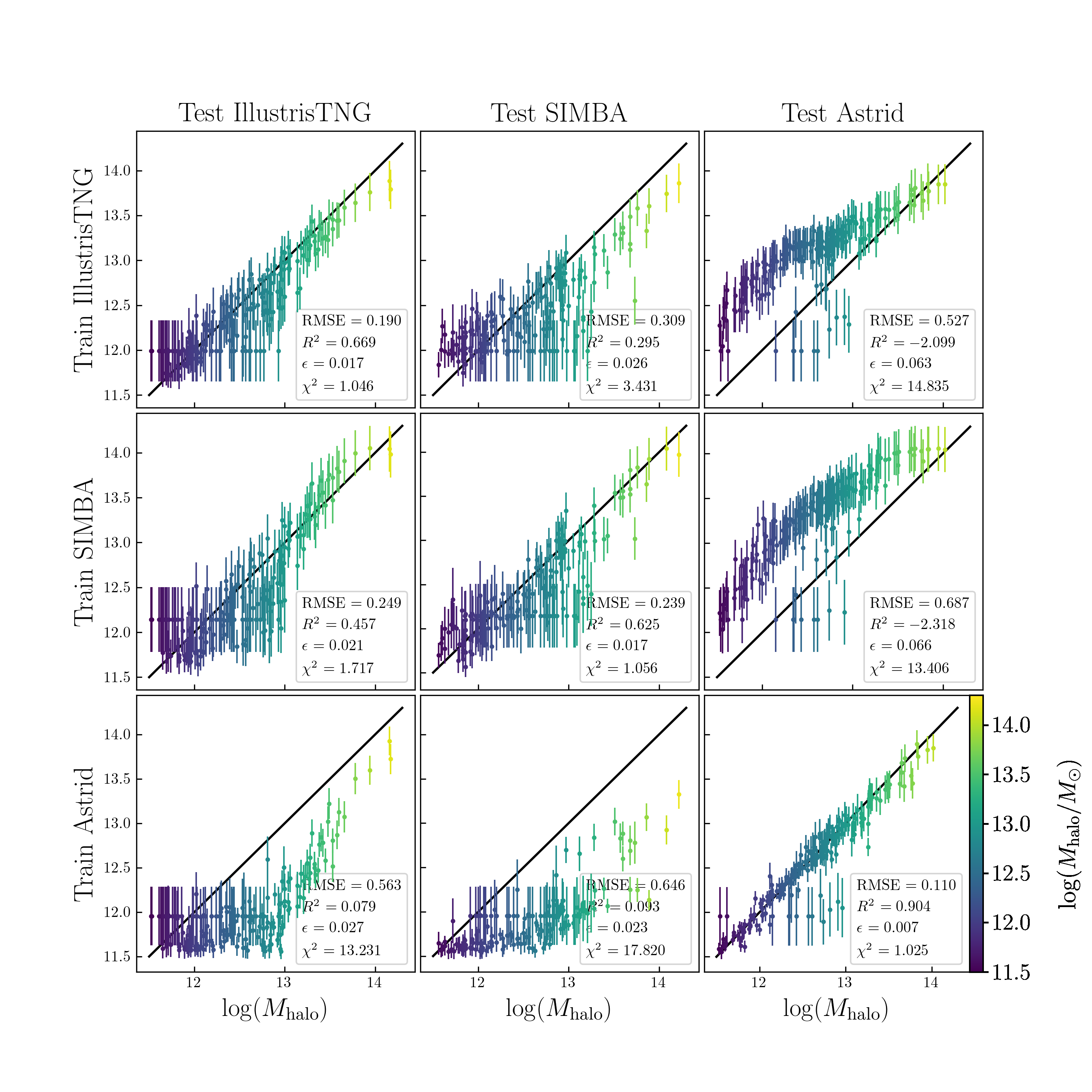}
    \caption{Cross-simulation results for IllustrisTNG, SIMBA, and Astrid on X-ray for $\Mhalo$, with observational limits. The x-axis of each panel corresponds to the true values of $\Mhalo$, and the y-axis corresponds to the inference values of $\Mhalo$, as before. The y-axis labels indicate that the panels in the top row were trained on IllustrisTNG, the middle row on SIMBA, and the bottom row by Astrid. The columns are labelled such that the panels in the first column were tested on IllustrisTNG, the second column's panels on SIMBA, and the third on Astrid. The diagonal panels are the result of training and testing on the same simulation. Training and testing on Astrid provide the tightest constraints and the best inference. These points are a fraction of the full dataset.}
    \label{fig:cross_sim_Xray_obs_Mhalo}
\end{figure*}

\begin{figure*}
    \centering
    \includegraphics[scale=0.7]{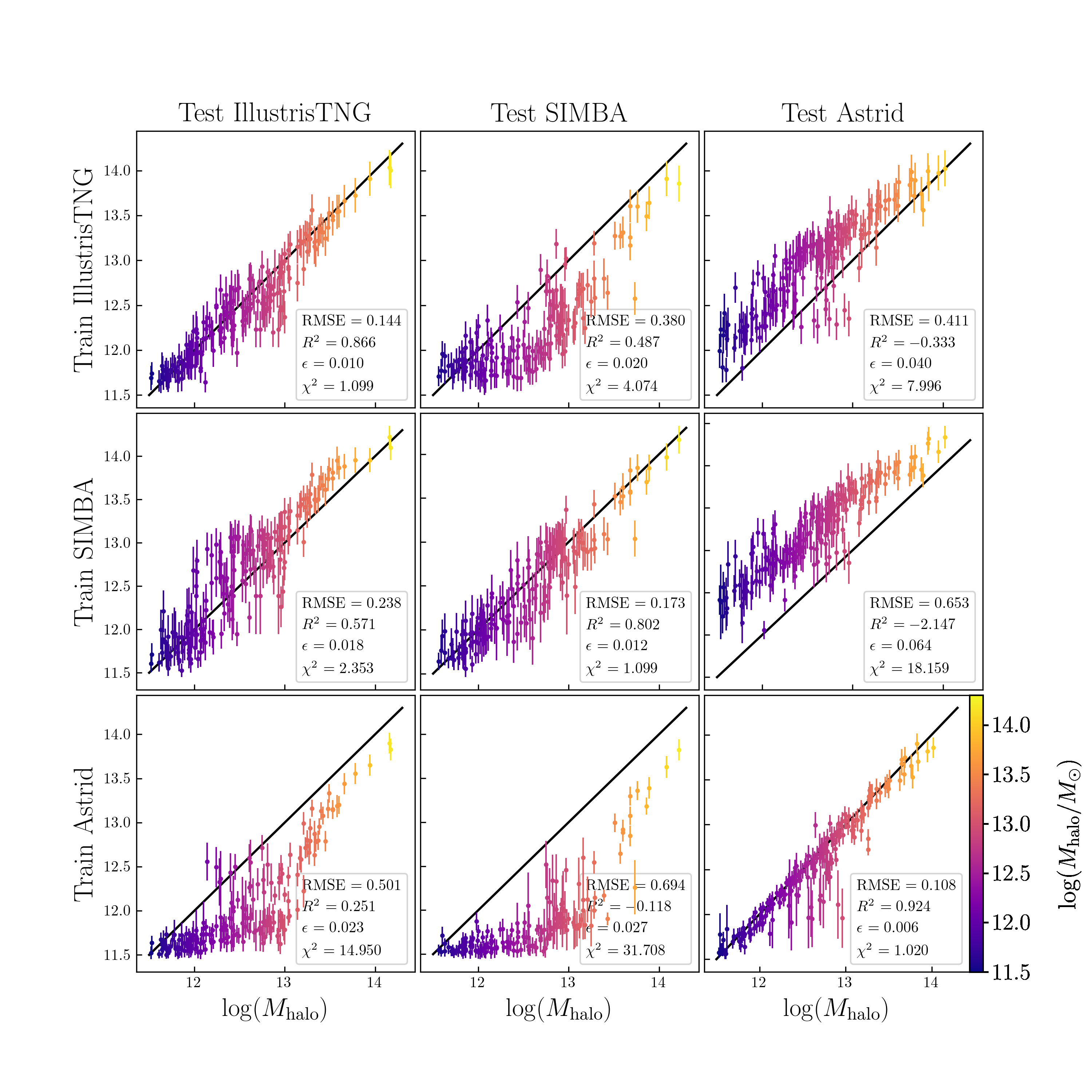}
    \caption{Cross-simulation results to infer $\Mhalo$ using the multifield with observational limits for IllustrisTNG, SIMBA, and Astrid. The layout is the same as in Fig.~\ref{fig:cross_sim_Xray_obs_Mhalo}. Even with the observational limits of HI and X-ray, training and testing on Astrid have the best overall inference for $\Mhalo$. }
    \label{fig:cross_sim_multi_Mhalo}
\end{figure*}

\begin{figure*}
    \centering
    \includegraphics[scale=0.7]{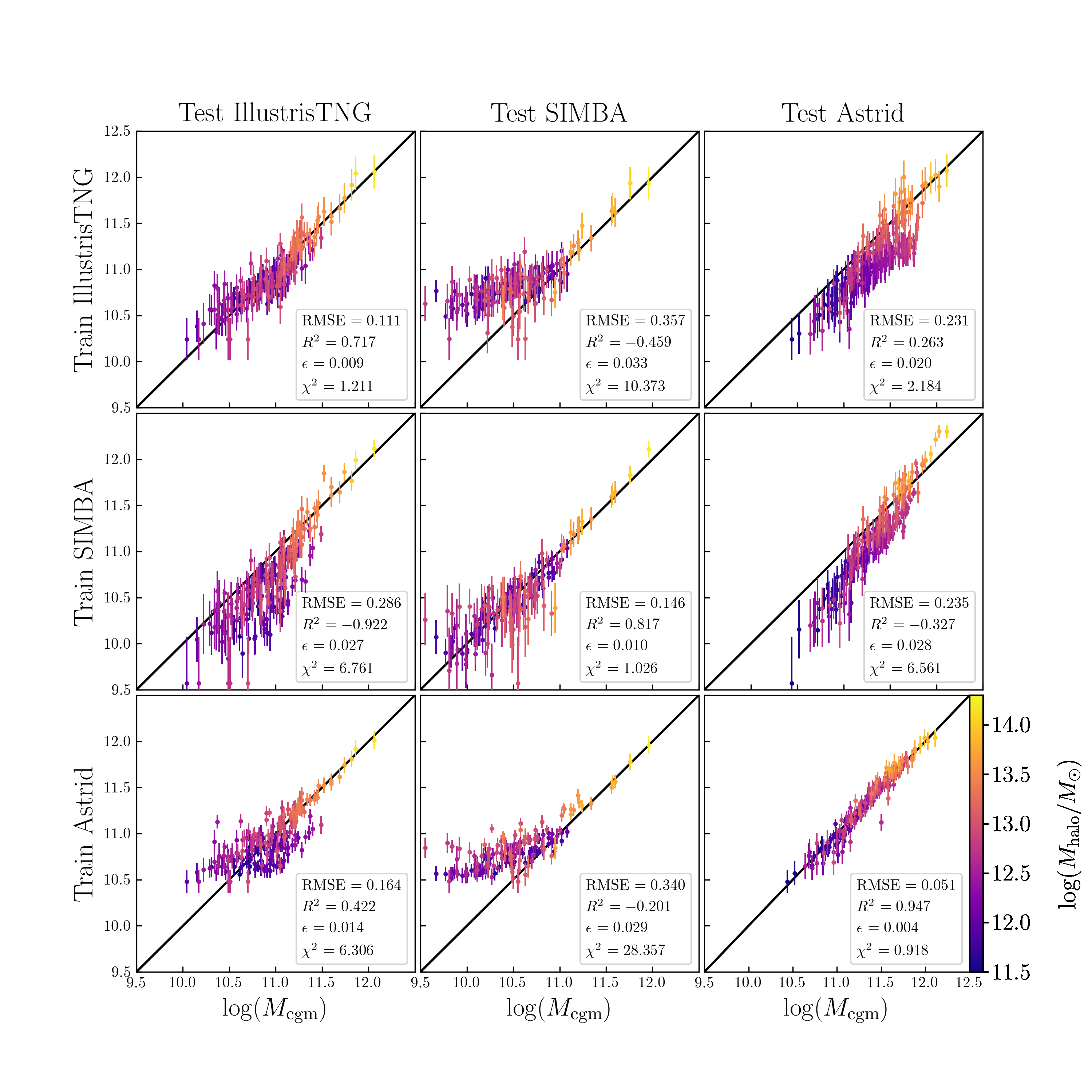}
    \caption{Cross-simulation results to infer $\Mcgm$ using the multifield with observational limits for IllustrisTNG, SIMBA, and Astrid. The layout is the same as in Figure~\ref{fig:cross_sim_Xray_obs_Mhalo}.}
    \label{fig:cross_sim_multi_Mcgm}
\end{figure*}

We present our main results in this section. First, we discuss training and testing on one idealised field at a time for the same simulation (single-simulation analysis), focusing on three inferred properties: 1) halo mass in $R_{\rm 200c}$ ($\Mhalo$ in \S\ref{sec:halo_mass}), 2) the mass ratio of CGM gas to the total mass inside $R_{\rm 200c}$ ($\fcgm$ in \S\ref{sec:fcgm}), and 3) the metallicity of the CGM inside 200~kpc ($\logZ$ in \S\ref{sec:logZ}). We do not display the case of a multifield here, as the results do not indicate a significant improvement. Following this, we show the results for the observationally limited case (for properties $\Mhalo$ and $\Mcgm$), strengthening the motivation for using a multifield (\S\ref{sec:obslimit}). We also organise network errors (RMSE) by mass bin (see Table~\ref{table:mass_bins}) and by simulation (training and testing on the same simulation) for the multifield case with observational limits (\S\ref{sec:cnn_errors}). Finally, we provide the results of a cross-simulation analysis encompassing all three simulations, with and without observational limits for comparison (\S\ref{sec:cross_inference}). 

We utilise Truth--Inference scatter plots to display inferences on the CGM properties. Each plot consists of multiple panels distinguished by field, simulation, or both. The panels visualise the true value, $\rm{TRUE}_i$, on the x-axis and the inferred posterior mean, $\mu_i$, on the y-axis, with error bars corresponding to the posterior standard deviation, $\sigma_i$. Four statistics (for the subset of data plotted and \textit{ not} the entire dataset) are also provided for each panel: the root-mean-square-error (RMSE, Equation~\ref{eqn:rmse}), coefficient of determination values ($R^2$, Equation~\ref{eqn:r_squared}), mean relative error ($\epsilon$, Equation~\ref{eqn:epsilon}), and reduced $\chi^2$ values (Equation~\ref{eqn:chi2}). The definitions and equations for each are given in \S\ref{sec:network_outputs}. The black diagonal line also represents a ``perfect inference'' one-to-one line between the true and inferred values.

\subsection{Inferred Halo Mass}
\label{sec:halo_mass}

The halo mass emerges as a readily interpretable property, directly deducible from the network, owing to its clear expectations: ``true'' high-mass halos should yield correspondingly high ``inferred'' halo masses, regardless of the simulation used for training and testing. Figure~\ref{fig:Mhalo_idealized_fields} illustrates the Inference--Truth plots for $\Mhalo$ across all three simulations for a subset of the data. We define the halo mass in Equation~\ref{eqn:mhalo} as the sum of dark matter, gas, and stars within $r<R_{\rm 200c}$.

The top row corresponds to the results using idealised X-ray maps to infer $\Mhalo$, and similarly for the bottom row using HI maps. The columns are ordered by the simulation used for training and testing: IllustrisTNG (left), SIMBA (middle), and Astrid (right). The points are coloured by halo mass throughout. We examine the CNN with input X-ray maps first. In the first panel, we train and test on IllustrisTNG and obtain inferences that indicate a relatively well-constrained monotonic relationship. Some points remain further away from the ``truth'' values set by the black diagonal line at the low-mass end, suggesting that X-ray may not be the best probe for these low-mass halos. The next panel visualises the training and testing results on SIMBA, with a slight improvement in the higher mass range and an overall relatively well-constrained, monotonic trend with a few outliers that stray far from the black line. This is exactly what is expected and is the same across simulations and fields. There are a few more outliers than IllustrisTNG and slightly larger error bars across the entire mass range. Finally, the third panel demonstrates that the CNN trained on and tested with Astrid has excellent inference power, as indicated by smaller error bars throughout the mass range.  

We can now look at the results obtained using the HI input maps. In the first panel, with training and testing on IllustrisTNG, we obtain a clear and well-constrained monotonic relationship with relatively little scatter. There is a slight improvement in error predictions when using HI instead of X-ray, as indicated by the change in the $\chi^2$ value from 0.918 with X-ray to 0.938 with HI. Visually, we can see the improvement in the lower mass range, as there is less scatter. The middle panel shows training and testing on SIMBA, where the inference made is significantly worse than it is with X-ray throughout the entire parameter space, especially intermediate to low masses with increased scatter and larger error bars. The last panel shows training and testing on Astrid. Training and testing the CNN on Astrid with HI input maps yields the best inference of the three galaxy formation models, indicated by the highest $R^2$ and lowest $\epsilon$. However, the inference made with HI is outperformed by that made with X-ray in the intermediate to high mass range.

Quantitatively, Astrid provides the most accurate and precise inference for both fields following the RMSE and $\epsilon$ values, respectively. It also has the highest $R^2$ score, indicating that a CNN trained and tested on Astrid can best explain the variability in the data. SIMBA has the lowest $R^2$ value overall with HI input maps, making it the least accurate in this case. Investigating the $\chi^2$ values, CNNs training and testing on 1) IllustrisTNG consistently overestimate the error ($\chi^2 < 1$), 2) SIMBA consistently underestimate the error ($\chi^2 > 1$), and 3) Astrid overestimate the X-ray error to a greater degree than IllustrisTNG but slightly underestimate the HI error to a lesser degree than SIMBA.

To interpret the meaning of the $\chi^2$ values reported in Fig. \ref{fig:Mhalo_idealized_fields}, we determine the percentage of $\sigma_i$ errors of individual data points that overlap the line of perfect inference. If the errors are truly capturing the Gaussian behaviour, as in $\chi^2=1$, we would expect $1-\sigma$ or 68\% errors to overlap. Briefly, we find that the percentage of overlapping points is 78. 3\% for the overall lowest $\chi^2=0.858$ for Astrid on idealised X-ray maps, and 65.0\% for the overall highest $\chi^2=1.277$ for SIMBA for idealised HI maps. In the case of SIMBA X-ray, we find an overlapping percentage of 68.9\% for $\chi^2=1.125$, which indicates a slight non-Gaussian behaviour for a $\chi^2$ just under one. Note that data points with underestimated errors generally overcontribute, especially this $\chi^2$ value. However, it is encouraging to see that the $\chi^2$ values scale as expected, meaning that they have a diagnostic value, but that the inferred errors do not completely follow Gaussian statistics.

\subsection{Inferred CGM Gas Fraction}
\label{sec:fcgm}

Figure~\ref{fig:fcgm_idealized_fields} shows the Truth--Inference plots for $\fcgm$ in the same format as Fig.~\ref{fig:Mhalo_idealized_fields}, with the colour bar still indicating $\Mhalo$. We see that $\fcgm$ does not have a monotonic trend, seen explicitly in Fig.~\ref{fig:counts}. Higher masses tend to be more constrained, illustrated by a less deviation from the black line and smaller errors than those of lower mass halos. However, this is likely due to having fewer higher mass halos for the network to learn from. We define $\fcgm$ in Equation~\ref{eqn:fcgm}, as the sum of non-star-forming gas within a radius of $r<200\ \rm{kpc}$ divided by the halo mass.

 CNN performs poorly with IllustrisTNG on idealised X-ray maps, resulting in scattered points with large error bars. The network underestimates the error bars, as indicated by a $\chi^2$ value greater than one. The next panel shows the results with SIMBA, for which there is better agreement and less scatter toward the higher and intermediate halo masses. However, for the low-mass halos, there is no distinctive trend, though the network can predict the values well overall but with somewhat large error bars (also underestimated). SIMBA also has slightly lower $\fcgm$ values than IllustrisTNG (c.f., Fig.~\ref{fig:counts}). Finally, a CNN with Astrid provides excellent inference for $\fcgm$ and accurately estimates the network error. The values are systematically larger, matching Fig.~\ref{fig:counts}.

Similarly, we display HI in the bottom row, with overall trends matching those seen with X-ray. However, HI offers tighter constraints at lower mass halos (higher $\fcgm$ values). This indicates that HI is a slightly better probe for $\fcgm$ than X-ray. Interestingly, SIMBA now overestimates the network error ($\chi^2$ value less than one), while IllustrisTNG and Astrid underestimate the errors. In general, $\fcgm$ performs worse than $\Mhalo$, but CNN is learning to infer this property using a single idealised field. 

It does not appear that the quality of inference by the CNN depends on where the range of $\fcgm$ lies with respect to the entire value space spanned by all three simulations -- IllustrisTNG returns the worst performance but has intermediate $\fcgm$ values, with an underestimate of the error. Astrid yields the most accurate and precise inferences for X-ray and HI fields, with lower scatter and error values for predicting $\Mhalo$ compared to IllustrisTNG and SIMBA. While SIMBA generally performs worse, it exhibits relatively good results in this case, especially with HI.

Following the results of \cite{davies_2020} using IllustrisTNG-100, we see similar non-monotonic trends using CAMELS-IllustrisTNG in $\fcgm$ as a function of halo mass. Low mass halos ($\log(\Mhalo/M_{\odot})< 12$) show high $\fcgm$ values. When the halo mass is slightly increased, there is a decline in the values of $\fcgm$, until approximately $\log(\Mhalo/M_{\odot}) \approx 12.5$ as a threshold mass, after which the monotonic trend with the halo mass returns. Star-forming feedback processes below this threshold mass are dominant and incapable of clearing the CGM. At the threshold mass, these star-forming feedback processes become stronger. Instead of learning the CGM of its gas, the AGN feedback is shut down as early black hole formation is limited \citep{delgado_2023}. We then see a dramatic increase due to turning on jet-mode feedback. Even for cluster-mass objects, AGN feedback cannot overcome deep potential wells, so that we again see high values of $\fcgm$. SIMBA has the strongest feedback implementation of the galaxy formation models considered, resulting in lower overall $\fcgm$ values throughout the entire mass range. Additionally, we note that although SIMBA has the largest scatter in $\fcgm$, this is not simply a reflection of larger statistical fluctuations, as it has a comparable amount of sub-L* objects to IllustrisTNG (see Table~\ref{table:mass_bins}). Astrid has the weakest feedback, resulting in higher overall $\fcgm$ values across the mass range \cite{ni_camels_2023}. Further analysis is needed to more concretely establish the relationships between feedback and halo mass such that these results are robust to observational data.

\subsection{Inferred Metallicity}
\label{sec:logZ}

Figure~\ref{fig:logZ_idealized_fields} shows the Truth--Inference plots for metallicity, plotted as the logarithm of the absolute value of $Z$ (note $\log(Z_{\odot})=1.87$, \citet{asplund_2009} on this scale. Metallicity presents an interesting challenge to our CNN, as there are often $\sim 1$~dex of scatter in $Z$ at the same halo mass with no obvious trend (see Fig.~\ref{fig:counts}). When training and testing on IllustrisTNG (top left), we see that higher-mass halos are slightly better constrained than low-mass halos, which are more scattered and have larger (and overall underestimated) error bars. We define the metallicity of the CGM in Equation~\ref{eqn:Zcgm} as the sum of the metallicity of the gas particles times the mass of non-star-forming gas within a radius of $r<200 \rm{kpc}$.

Training and testing on SIMBA results in significant scatter across the entire mass range, with larger and underestimated error bars. $L^*$ and group halos have higher metallicity values overall than in the previous panel. The last panel shows training and testing with Astrid, returning the best overall inference in $\logZ$ across the entire mass range. Although the error is underestimated, Astrid has much higher accuracy and precision based on RMSE, $R^2$, and $\epsilon$ values. We argue that this is quite an impressive demonstration of our CNN's ability to predict a value with significant scatter at a single halo mass.  

The bottom row illustrates this same inference, but now using HI, where we see similar trends as with X-ray, though slightly more constrained in the case of IllustrisTNG and SIMBA and slightly less constrained in the case of the low mass end of Astrid. The same upward shift for $L^*$ and group halos is seen with SIMBA, alluding to SIMBA's strong astrophysical feedback prescriptions that impact higher mass halos. This is also seen in the changes in lower (higher) $\chi^2$ values for IllustrisTNG and SIMBA (Astrid). We conclude that neither X-ray nor HI is powerful enough on its own to infer $\logZ$. Surprisingly, the entire metallicity of the CGM can be well inferred using HI, especially in the case of Astrid, despite being a small fraction of overall hydrogen, which itself is a primordial element. We do not attempt to provide a physical interpretation of the metallicity of the CGM, as it is quite complex and will be a good topic to focus on for our future work applying interpretative deep learning techniques.

\subsection{Observational Limits and Multifield Constraints}
\label{sec:obslimit}

Simulations must consider the limitations of current and future observational multiwavelength surveys, such that a one-to-one correlation between them and the developing models can exist. The specific limits used in this work come from the eROSITA eRASS:4 X-ray luminosity of $2\times 10^{-13}$ erg s$^{-1}$ cm$^{-2}$ arcmin$^{-2}$, and the typical radio telescope column density for measurements of HI as $10^{19}$ cm$^{-2}$. Again, we include the RMSE values $R^2$, $\epsilon$, and $\chi^2$, which are especially important to distinguish between single and multifield inferences.

The top row of Figure~\ref{fig:TNG_all_obslimits} displays the Truth--Inference plots, highlighting the power of using multiple fields to infer $\Mhalo$ by training and testing a CNN with the IllustrisTNG observationally limited datasets. Utilising the X-ray (top left), it is clear that we cannot make an inference toward lower halo masses (Sub-L*). This is expected, given the {\it eROSITA}-inspired limits, which show X-ray emission strongly correlating with the halo mass, following Fig.~\ref{fig:maps_fn_mass_v2}. The inability to make a clear inference in this mass regime despite providing the CNN with the most information (nearly 3500 separate Sub-L* halos, see Table~\ref{table:mass_bins}) reiterates the weaknesses of X-ray. The X-ray inference improves in the L* range, but is still highly scattered.  \citet{chadayammuri2022} targeted L* galaxies by stacking {\it eROSITA} halos and found a weak signal, which appears to be supported by the assessment here. The groups provide much better inference for $\Mhalo$ since these objects should be easily detectable via {\it eROSITA}. In the middle panel, we explore HI with observational limits to infer $\Mhalo$. Interestingly, HI does a much better job for sub-L* halos, as these are robustly detected in the 21-cm mapping (see Fig.~\ref{fig:maps_fn_mass_v2}).  The inference worsens for L* halos and for much of the Group range. HI thus far shows improvements via a lower RMSE value, a $R^2$ value closer to 1, and a lower $\epsilon$ value. It also indicates that the network predicts a greater error underestimation due to a higher $\chi^2$ value.

As neither X-ray nor HI is robust enough to infer $\Mhalo$ alone properly, we now train and test the network on combined HI + X-ray ``multifield''. The multifield approach is specifically used when one field alone may not be enough to constrain a property fully or only constrain a property within a certain range of values. The secondary or tertiary fields would then be able to fill in some gaps or tighten the constraints within the inference. Additionally, with the ability to adjust the network based on current observations, we form computational counterparts to future surveys to aid in its construction. We achieve stronger constraints throughout the entire mass range, even with observational limits from both X-ray and HI. X-ray probes the L* and Group mass range well, while HI probes the sub-L* mass well, alleviating the previously unresolved noise of the left panel. We see a quantitative improvement in the multifield approach in lower values for RMSE, $R^2$, and $\epsilon$, which comes at the price of increased underestimation of errors seen with a slight increase in $\chi^2$ value. 

The bottom row of Fig.~\ref{fig:TNG_all_obslimits} provides similar results and trends for $\Mcgm$ (defined in Equation~\ref{eqn:Mcgm}) via IllustrisTNG with X-ray, HI, and multifield using observational limits. X-ray here is also not powerful enough as a probe to infer this property, especially in the low halo mass region. We then look at HI, where there is a better overall inference in the low halo-mass region. However, HI produces more scatter towards the high halo masses than X-ray. The last panel displays results from the HI+X-ray multifield, which is an overall improvement compared to either field alone. The constraints are tighter overall, and the scatter is reduced, as seen in the RMSE values, $R^2$, $\epsilon$, and $\chi^2$. Additional Truth--Inference multifield plots with observational limits for the remaining CGM properties can be found in the Appendix~\ref{sec:appendix_more_truth_inf_plots}.

\subsubsection{Visualizing the CNN Error}
\label{sec:cnn_errors}

To quantify the CNN error across all six CGM properties ($\Mhalo$, $\fcgm$, $\logZ$, $\Mcgm$, $\fcool$, and $\logT$), we plot the error in each property binned by the halo mass. In the left panel of Figure~\ref{fig:err_obslimit_type}, we plot the error (neural network error, or mean relative error) for each property when considering the observational limits on HI, X-ray, and multi-field HI + X-ray for a CNN that is trained and tested on IllustrisTNG. Panels are separated by halo mass, where we use the full dataset instead of the subset in the Truth--Inference plots.

We outline the general trends of this figure and point out interesting features. In the Sub-L* panel, X-ray maps alone provide the highest error, followed by multifield, and then HI with the lowest error, to be expected. Note that there is an infinitesimal difference between multifield and using HI alone. In the second panel (L*), the margin of error between X-ray and HI is decreasing, meaning that X-ray is becoming increasingly more important in the intermediate halo mass range. Multi-field development is also strictly improving with the use of HI alone. With Groups, the multifield offers a greater improvement over either field alone, except for $\log(\Mcgm)$ where the X-ray has a slightly lower error. 

Focusing on $\fcgm$, the errors are generally smaller than those of $\Mhalo$, but this may reflect the quantity range that is inferred as $\fcgm$ is mainly between $0.0-0.16$ while $\Mhalo$ varies between $11.5-14.3$. Meanwhile, $\logZ$ has similar error levels between HI and X-ray, with a small improvement for multifield for sub-L * and L *. The errors in $\logZ$ vary between $0.16-0.24$, so measuring metallicity at this level of accuracy is promising, but distinguishing high values of metallicity from low ones is disappointing for IllustrisTNG (see Fig.~\ref{fig:c1x3panel_logZ}).  

The last three sets of properties, $\log(M_{\rm cgm})$, $\fcool$, and $\log(T)$, have not been previously illustrated as Truth--Inference plots. They depict similar trends and show general multifield improvement. HI infers sub-L* the best, while X-ray infers Groups the best. The multifield is most important for L* halos, and across all six properties, there is a significant improvement in the inference. Other halo categories usually do not result in as much improvement; in some cases, the multifield performs slightly worse.  We note that inference of $\fcool$ for groups is a significant improvement, from $0.102$ (X-ray) and $0.125$ (HI) to $0.084$ (multifield), reflecting that CNN integrates observations of both cool gas (HI) and hot gas (X-ray) in this fraction.

The right panel of Fig.~\ref{fig:err_obslimit_type} outlines the errors in IllustrisTNG, SIMBA, and Astrid for the multifield HI+X-ray with observational limits for all six properties. The halo mass again separates the three panels. Generally, a CNN trained and tested on SIMBA has the highest error over the entire mass range, while a CNN trained and tested on Astrid returns a better inference. $\fcgm$ breaks this trend, as it is significantly worse for L* mass halos when using IllustrisTNG, which is directly due to the drastic inflection point seen in Fig.~\ref{fig:counts}. Additionally, Astrid can infer $\logZ$ remarkably well for L* mass halos, compared to the high error when using IllustrisTNG. This can be seen in Fig.~\ref{fig:c1x3panel_logZ} where IllustrisTNG has much more scatter across the entire mass range, while Astrid shows little scatter.

\subsection{Cross Simulation Inference}
\label{sec:cross_inference}

Until now, each Truth--Inference plot has been created by training and testing on the same simulation. In this section, we provide the results obtained when training on one simulation or galaxy formation model and testing on another to prove the degree of robustness across any particular simulation. We do this for both an X-ray with observational limits and a multifield with observational limits.

In Figure~\ref{fig:cross_sim_Xray_obs_Mhalo}, we demonstrate the cross-simulation inference between IllustrisTNG, SIMBA, and Astrid, using X-ray with observational limits only on the $\Mhalo$ property. The diagonal plots correspond to the training and testing in IllustrisTNG, SIMBA, and Astrid from upper left to lower right (repeated from the upper panels of Fig.~\ref{fig:Mhalo_idealized_fields}).

The top row refers to CNNs trained on IllustrisTNG, where each panel from left to right has been tested on IllustrisTNG, SIMBA, and Astrid, respectively. When tested on SIMBA, most points are close to the black line, but with significantly more scatter. When tested on Astrid, we can only recover good constraints for the high-halo mass range. There is much more scatter in the low-mass range, as a majority of them are overestimated, except for a few outliers, most likely resulting from the inability of X-rays to probe the low-halo mass range. 

When training on SIMBA, but then testing on IllustrisTNG, there is still quite a bit of scatter in the low-mass halos, and the high-mass halos are now overestimated. This matches the expectations from the brightness differences between IllustrisTNG (brighter) and SIMBA (dimmer). When testing on Astrid, all points are shifted up and overestimate halo mass. 

Finally, training on Astrid and testing on IllustrisTNG cannot recover any of the results. There is a lot of scatter for the low-halo-mass range with large error bars, with points that do not follow the expected trends in IllustrisTNG for intermediate and high masses. Astrid underestimates the majority of the halo masses. When testing on SIMBA, the results cannot be recovered either, as most points underestimate the halo mass. Although training and testing on Astrid seem to provide the best constraints on halo mass with X-ray observational limits, it is the least robust simulation out of the three, as measured by its ability to be applied to other simulations as a training set. In contrast, other models trained on the Astrid LH set \citep{ni_camels_2023, di_santi_2023} are the most robust, as the parameter variations produce the widest variation in galaxy properties, in turn making ML models more robust to changes in baryonic physics. IllustrisTNG is the most robust in this case, as it returns the results of the other two simulations with the least amount of scatter.

One oddity in the statistical measurements produced comes from training on either IllustrisTNG or SIMBA and testing on Astrid, which results in a negative $R^2$ value, indicating a significant mismatch in the models. Another unusual statistic is in the extremely high $\chi^2$ values from three cases: 1) training on IllustrisTNG and testing on Astrid, 2) training on SIMBA and testing on Astrid, and 3) training on Astrid and testing on either IllustrisTNG or SIMBA. Each reiterates the lack of robust results that can be achieved with Astrid.

Figure~\ref{fig:cross_sim_multi_Mhalo} illustrates the cross-simulation results on $\Mhalo$ with observational limits on the multifield HI+X-ray for IllustrisTNG, SIMBA, and Astrid. The top left panel shows this multifield, trained on and tested with IllustrisTNG, where overall, $\Mhalo$ can somewhat be constrained throughout the entire parameter space. The second panel on the diagonal corresponds to the same multifield but is now trained on and tested with SIMBA. The constraints here are weaker throughout the entire parameter space as there is more overall scatter, though the trend is the same as expected. The last panel on the diagonal shows the network trained on and tested with Astrid, where we can obtain the tightest constraints overall, especially in the higher halo mass range. The few outliers towards the mid (L*) to low (Sub-L*) mass range with larger error bars may need further investigation. 

The top row shows training with IllustrisTNG and testing on IllustrisTNG, SIMBA, and Astrid, respectively. When training on IllustrisTNG and testing on SIMBA, we expect that for a given mass halo in IllustrisTNG, that same halo will look dimmer and, therefore, less massive in SIMBA. This is seen here, as most halos are below the black line. When the network is now tested on Astrid, a similar but opposite expectation is met. With the knowledge that for a given halo mass in IllustrisTNG, that same halo will look brighter and, therefore, more massive in Astrid, this trend also makes sense, as we see a large majority of the points shifted above the black line. We can conclude that with observational limits in the multifield, training on IllustrisTNG can return the trends in SIMBA and Astrid, but there is an offset in recovered $\Mhalo$ explainable by the shift in observables.

The middle row shows training with SIMBA and testing on IllustrisTNG, SIMBA, and Astrid, respectively. When the network trains on IllustrisTNG, it can recover the inference and achieve good constraints. The same halo in SIMBA will appear brighter in IllustrisTNG, so the shift in most points upward above the black line is, therefore, as expected. When testing on Astrid, we still recover the inference and achieve good constraints, but we see the same shift as we saw when training on IllustrisTNG and testing on Astrid. This also aligns with the expectations, as the halos in Astrid will seem much brighter than those in SIMBA. We can conclude that with observational limits in the multifield, SIMBA is also robust enough to recover inference and constraints for $\Mhalo$.

The bottom row shows training with Astrid and testing with IllustrisTNG, SIMBA, and Astrid, respectively. When the network tests on IllustrisTNG, we can recover the general trend with slightly less strong constraints. We can recover the general trend with slightly less strong constraints when the network is tested on SIMBA. The halos in Astrid will be brighter than the same halos in IllustrisTNG and SIMBA, so the majority of the points are below the black line when testing on IllustrisTNG and SIMBA. We can conclude that a CNN trained on Astrid cannot recover the inference and constraints for $\Mhalo$. We see the same statistical nuances as in the previous figure: negative $R^2$ values and large $\chi^2$ values in the same configurations.

By adding observational constraints for both HI and X-rays, the simulations gain a further level of similarity, which enhances their constraining power in the cross-simulation analysis. Figure~\ref{fig:cross_sim_multi_Mcgm} shows the results of using the multifield (HI+X-ray) approach with observational limits on $\Mcgm$, with observational limits. The layout of the plot is analogous to that of Fig.~\ref{fig:cross_sim_multi_Mhalo}. Training on IllustrisTNG (top row) overpredicts the results for intermediate and low mass halos when testing on SIMBA and underpredicts the same results when testing on Astrid. This aligns with the expectations in the bottom left panel of Figure~\ref{fig:counts}, which describes the relationship between the halo mass and $\Mcgm$. Training on SIMBA (middle row) underpredicts intermediate and low mass halos results when testing on IllustrisTNG and Astrid. Note that there is much more scatter when testing on IllustrisTNG, especially for objects with low $\Mcgm$ values. Training on Astrid (bottom row) does reasonably well when testing on IllustrisTNG with some scatter in the intermediate and low-mass halos. However, it overpredicts these intermediate and low-mass halos when tested on SIMBA. 

Although able to return similar trends, cross-simulation training and testing display offsets related to different CGM properties in all simulations. However, it is enlightening to see that cross-simulation inference improves when more bands are included, which indicates that broad properties like $\Mhalo$ and $\Mcgm$ are more robustly characterised by observing in multiple bands. We make a deliberate choice to show the cross-simulation analysis results for $\Mhalo$ and $\Mcgm$, not $\fcgm$, as it is a ratio of the gas mass to the halo mass throughout the halo (not within 200 kpc), leading to a more complex trend that is not as easily interpretable. Cross-simulation analysis can offer a way to understand the direction and magnitude of systematic offsets and the variations between feedback implementations qualitatively and the feedback energy as a function of redshift. This is entirely contingent on our ability to create physically motivated deep-learning models that are interpretable, which is the focus of our future work.

\section{Discussions}
\label{sec:discussion}

In this section, we discuss the interpretation of cross-simulation analysis (\S \ref{sec:cross_sec_interpret}), assess the applications and limitations of CNNs when applied to CGM (\S \ref{sec:limitations}), compare the variance between true and inferred values for $\log(\Mhalo)$, $\log(\Mcgm)$, and $\logZ$ using the idealized multifield maps (\S \ref{sec:compare_variance}), and expand on an intriguing direction for future work (\S\ref{sec:future_work}).

\subsection{Cross-Simulation Interpretability}
\label{sec:cross_sec_interpret}

In Section~\ref{sec:cross_inference}, we explore the robustness of simulations by examining cross-simulation inference with and without observational limits. Fig.~\ref{fig:cross_sim_multi_Mhalo} presents cross-simulation inferences for multifield HI + X-ray with observational limits on $\Mhalo$. Upon initial inspection, training, and testing on Astrid offer the tightest constraints across the entire mass bin. In general, a test simulation will overpredict (underpredict) properties when trained on a simulation with CGM observables that are dimmer (brighter). Among the three simulations, a CNN trained on IllustrisTNG is the most robust, as it accurately captures the differences between halo mass measurements when trained on SIMBA and Astrid. However, more work must be done to show that a CNN trained on IllustrisTNG will produce the most robust predictions when applied to real observational data. A novel aspect to further explore is training and testing on multiple simulations, varying the feedback parameters such that the CNN would marginalise over the uncertainties in baryonic physics.

The effort to train and test on different simulations mimics training on a simulation and predicting real observational data. Although it is disappointing to see such deviations in the results of the cross-simulation analysis, we know that some simulations offer better representations given the specific scope of this work than others. Using observational limits that resemble the ranges of detection of current instruments as a simulation constraint, we can begin directly comparing simulations and observations. We note that the simulations are unconstrained by available observations in the CGM. The fiducial prescriptions for IllustrisTNG and SIMBA are calibrated to match the available data of the groups with varying success \citep{oppenheimer_simulating_2021}, but Astrid with its higher $f_{\rm gas}$ values has not been calibrated similarly. Importantly, no simulation is a perfect representation of the real universe, but it is crucial to develop CNNs that can adapt to a wide range of mock halos generated using multiple galaxy formation codes that aim to simulate these systems with realistic physical prescriptions.

Robustness quantification, or how well a network trained on one simulation can infer a given quantity when tested on another simulation within any set of simulations and machine learning algorithms, including the CAMELS suites, is crucial to further their development \citep{villaescusa-navarro_robust_2021,villanueva_robustness_2022,Echeverri_2023,di_santi_2023}. The lack of robustness can be due to either 1) differences between simulations, 2) networks learning from numerical effects or artefacts, or 3) lack of overlapping between simulations in the high-dimensional parameter space. These reasons are not surprising, because of the use of the CV set within CAMELS, and there could be slight variations in feedback that are unaccounted for. Using the LH set instead would improve the results obtained in this work. Additionally, precision (smaller error bars) without accuracy (recovering the ``true'' values) is meaningless. Therefore, although Astrid generally has the smallest error bars, this alone shows strong biases when tested on other models. Future work can be done to address the inability to obtain robust constraints while performing cross-simulation analysis. One avenue is through domain adaptation \citep{Ganin_domain_adapt_2015}, which allows a smoother transition between training and testing on different simulations such that we obtain robust results.

\subsection{Applicability and Limitations of CNNs Applied to the CGM}
\label{sec:limitations}

We have applied a CNN following the structural format used by \citet{camels_2021_fundamental_params} and modified it to infer underlying {\it properties} of the CGM of individual halos with fixed cosmology and astrophysics within the CAMELS CV set. The former CNN infers six independent parameters (two cosmological and four astrophysical feedback) {\it by the design of the LH simulation set}. Our trained CGM CNN learns to predict properties with high co-dependencies (e.g., $\log(M_{\rm halo})$ and $\logT$) and related quantities ($\fcgm$ and $\Mcgm$). In the latter case, there are two different ways to quantify CGM mass in two distinct apertures-- $\Mcgm$ is the CGM mass inside 200~kpc, and $\fcgm$ is the mass of CGM over the total mass inside $R_{\rm 200c}$.

We attempted to infer one property at a time instead of all six and found only a marginal improvement.  CNN implemented in this work, classified as a moment network \citep{jeffrey2020}, has the flexibility to infer multiple properties simultaneously, but requires a rigorous hyperparameter search, as detailed in \S\ref{sec:cnn}. 

A concern that often appears with any simulation-based approach is the possibility of biases seeping into the result, generally due to incomplete modelling of physical processes. We aim to alleviate this concern first by using the CV set within the CAMELS simulations, where the values of cosmological and astrophysical feedback parameters are fixed to their fiducial values. The LH set, which was not used in this work (but could easily be integrated as part of future efforts), increases the chances of successful cross-simulation analysis as the astrophysical dependencies are completely marginalised. From this standpoint, the CV set is not best suited to produce robust cross-simulation analysis. Using the CV set, we gain valuable insight into the distinctions among simulations and their effects on the results of the CGM properties in this study. In addition to using the LH set, we can explore training and testing on more than one simulation or performing a similar analysis on the broader parameter space of TNG-SB28 \citep{ni_camels_2023}.

We apply CNNs to the CGM datasets to 1) determine the degree to which physical properties of the CGM can be inferred given a combination of fields and simulations, and 2) examine different observing strategies to determine how combining different wavebands can infer underlying CGM properties. 

We demonstrate the feasibility of applying a CNN to observational datasets and return values and errors for the CGM properties, including $\Mhalo$ and $\Mcgm$. Additionally, the inference of $\Mhalo$ is more robustly determined when another field, along with its associated observational limits, is added. However, training on one simulation and testing on another support the notion that predictions can produce significantly divergent results compared to the true values, as seen in Figs. \ref{fig:cross_sim_Xray_obs_Mhalo} and \ref{fig:cross_sim_multi_Mhalo}. As mentioned in \S\ref{sec:cross_sec_interpret}, although IllustrisTNG, SIMBA, and Astrid have been tuned to reproduce galaxies' and some gas properties, they make varied predictions for gaseous halos. In future efforts to improve this work, the LH set would replace the CV set, under the expectation of improvement, as all astrophysics is marginalised. Should this not be the case, domain adaptation is the longer-term solution to help bridge the many gaps between different subgrid physics models. Another interesting future direction would include training and testing on combinations of simulations, though this is ideally performed with the LH set.

\subsection{Multifield Variance Comparison}
\label{sec:compare_variance}

As an additional test, we check if our CNN-inferred values can reproduce the original dispersion of a CGM dataset. Even if a CNN can reproduce the mean value of a CGM parameter, can it also reproduce the spread of values? Fig.~\ref{fig:counts} shows the shaded $\pm 1\sigma$ dispersions in addition to the medians. We, therefore, calculate the dispersion for $\log(\Mcgm)$ and $\logZ$ for sub-L* and L* galaxies across the three simulations to explore our CNN's ability to reproduce this scatter in relatively flat $M_{\rm halo}$ bins. The values are displayed in Table~\ref{table:compare_variance}. On a positive note, it does appear that sub-L* and L* dispersions for $\log(\Mcgm)$ are well reproduced in SIMBA and Astrid. However, the dispersions are severely underestimated, often by a factor of two, for $\logZ$ and log($\Mcgm$), but, notably, $R^2$ measures and the performance of the CNN is poor for these cases. In particular, with IllustrisTNG, $M_{\rm CGM}$, and $f_{\rm CGM}$ as a closely related quantity, show worse performance due to rapidly changing gas fractions in response to feedback, as we discuss in \S\ref{sec:fcgm}. In this case, the CNN is unable to adequately learn signatures of reduced CGM mass at a fixed halo mass. This test presents a crucial challenge for future machine learning and deep learning methods in reproducing the spread of a given property for objects that are otherwise alike.

\begin{table}
\centering
\caption{The variance of $\Mcgm$ and $\logZ$ compared between the input from CAMELS (truth) and the values from the idealized multifield (HI+X-ray) inference.}

\label{table:compare_variance}
  \begin{tabular}{c c c c c}
    \toprule
    \multirow{2}{*}{$\log(\Mcgm)$} &
      \multicolumn{2}{c}{sub-L*} &
      \multicolumn{2}{c}{L*}  \\
      & {True} & {Infer} & {True} & {Infer} \\
      \midrule
    IllustrisTNG & 0.024 & 0.013 & 0.127 & 0.074 \\
    SIMBA & 0.099 & 0.094 & 0.129 & 0.111 \\
    Astrid & 0.026 & 0.024 & 0.038 & 0.038 \\
    \toprule
    \multirow{2}{*}{$\logZ$} &
      \multicolumn{2}{c}{sub-L*} &
      \multicolumn{2}{c}{L*}  \\
      & {True} & {Infer} & {True} & {Infer} \\
      \midrule
    IllustrisTNG & 0.072 & 0.041 & 0.078 & 0.035 \\
    SIMBA & 0.086 & 0.060 & 0.108 & 0.052 \\
    Astrid & 0.053 & 0.033 & 0.042 & 0.035 \\
    \bottomrule
  \end{tabular}
\end{table}

\subsection{Future Work}
\label{sec:future_work}

In expanding the scope of this work to additional wavelengths in the future, we also aim to advance our understanding of where the CNN extracts important information from within a given map. We can use the information gained from this type of analysis, which has not been applied to CGM data before this work, to inform future observational surveys on how best to achieve the greatest scientific returns given wavelength, survey depth, and other specifications. Additionally, this type of analysis will be necessary to determine machine learning verification and validation. To achieve this, we hypothesise that moving towards higher resolution simulations, including IllustrisTNG-100, EAGLE, and others along with a more physically motivated deep-learning model, will have a significant impact across a wide range of scales, especially in the case of observational limits.

\section{Conclusions}
\label{sec:conclusions}

In this study, we use convolutional neural networks (CNNs) trained and tested on CAMELS simulations based on the IllustrisTNG, SIMBA, and Astrid galaxy formation models to infer six broad-scale properties of the circum-galactic medium (CGM). We focus on the halo mass, the CGM mass, the metallicity, the temperature, and the cool gas fraction. We simulate two observational fields, X-ray and 21-cm HI radio, which can represent the broad temperature range of the CGM.  We tested our CNN on datasets with and without (idealised) observational limits. Our key findings include the following.

\begin{enumerate}

    \item[1.] When training and testing the CNN on the same simulation:

    \begin{enumerate}

        \item By comparing all the CGM properties the CNN is trained to infer, it performs the best overall on $\Mhalo$ and $\Mcgm$, both with and without observational limits. For IllustrisTNG with observational limits, the RMSE values returned for $\Mhalo$ are $\sim 0.14$ dex, and $\Mcgm$ are $\sim 0.11$ dex when combining X-ray and HI data.  

        \item The ``multifield'' CNN trained simultaneously on X-ray and HI data with observational limits allows for the best inference across the entire mass range using the same inputs without the discontinuities seen when trained only on one field. Obtaining interpretable inferences on the halo mass for the continuous range of $11.5 \le \log(\Mhalo/M_{\odot}) \le 14.5$ requires a multifield, although various combinations may be better over smaller mass bins than others. Sub-$L^*$ halos ($\Mhalo=10^{11.5-12}\ {\rm M}_{\odot}$) are only marginally better inferred with HI than multi-field. Moving to $L^*$ halos ($\Mhalo =10^{12-13}\ {\rm M}_{\odot}$) and the more massive groups ($\Mhalo>10^{13}\ {\rm M}_{\odot}$), there is a drastic improvement when using multi-field over X-ray and HI alone. Our exploration demonstrates that CNN-fed multiple observational fields with detectable signals can continuously improve the inference of CGM properties over a large mass range given the same input maps. 
    
        \item When adding observational limits to the multifield CNN, the inference accuracy declines, but still returns RMSE values indicating success. Recovering total mass from observations appears to be feasible with our CNN. HI mapping is especially critical for recovering CGM properties of sub-L* and L* galaxies.

    \end{enumerate}

    \item[2.] For CNN cross-simulation analysis (training on one simulation and testing on another):

    \begin{enumerate}
    
        \item When applying cross-simulation analysis by training on one simulation and testing on another, the inferred values generally correlate with the true physical properties. Still, they are frequently offset, indicating strong biases and overall poor statistical performance.  
     
        \item Interestingly, the cross-simulation analysis reveals that using the HI+X-ray multifield with observational limits improves the halo mass inference compared to that from X-ray maps alone. In the process of adding constraints in this case, the difference between the individual simulation parameter spaces becomes smaller and acts as tighter boundary conditions for the network. 

    \end{enumerate}

\end{enumerate}

Our results have broader implications for applying deep learning algorithms to the CGM than those outlined here. First, performing a cross-simulation analysis and determining that the CNN is robust opens the possibility of replacing one of the simulations with real data to infer the actual physical properties of observed systems. Second, the addition of more wavelengths is easily implemented within image-based neural networks. To continue making connections to current and future multiwavelength surveys, we can expand the number of fields used in this architecture beyond X-ray and HI, including image-based CGM probes like the Dragonfly Telescope that can map the CGM in optical ions, like $\rm{H}\alpha$ and $\rm{NII}$ \citep{lokhorst_dragonfly_2022}, and UV emission from ground- or space-based probes \citep{johnson_MgII_2014, johnson_2018, burchett_2018, peroux_2020}. Most importantly, this method would allow simulation differences to be marginalised, while still obtaining correlations and constraints. We can overcome the current challenges of cross-simulation analysis by training our CNN on multiple CAMELS simulations and parameter variations existing and in production (including expanding to EAGLE \citep{schaye_eagle_2015}, RAMSES \citep{teyssier_ramses_2010}, Enzo \citep{brian_enzo_code_2014}, and Magneticum\footnote{\url{http://www.magneticum.org/}}) while integrating additional wavebands. It is crucial to identify the primary source of information for CNN to increase the number of simulations and wavelengths used as input. Future work includes performing saliency analysis with integrated gradients to determine the most important pixels on a given map. It allows for more targeted and efficient adjustments to improve inferences. This can reveal which underlying physical properties are universally recoverable and robustly predictable in observations.

The CGM demarcates a region of space defined by nebulous boundaries, which poses a unique challenge to traditional analysis techniques like Principal Component Analysis (PCA). In addition, there are no established methods to characteristically analyse CGM. The phrase ``characteristically analysing'' implies distinctly categorising entities. For instance, traditional analysis can be used with galaxies to classify them into various categories based on their unique evolutionary traits, as evidenced by \citet{Lotz_2004}. However, the CGM refers to the area surrounding the galactic disk until the accretion shock radius, where neither boundary is precisely defined as they cannot be directly observed. Applying the same traditional analysis approach to a CGM dataset would require a rigid pipeline, making it difficult to incorporate new simulations or wavelengths without extensive reconfiguration. Deep learning offers a more flexible and versatile approach as a solution.

\section*{Acknowledgements}
We thank Shy Genel and Matthew Ho for valuable feedback and suggestions for the paper. The CAMELS simulations were performed on the supercomputing facilities of the Flatiron Institute, which is supported by the Simons Foundation. This work is supported by the National Science Foundation (NSF) grant AST 2206055 and the Yale Center for Research Computing facilities and staff. The work of FVN is supported by the Simons Foundation. The CAMELS project is supported by the Simons Foundation and the NSF grant AST 2108078. DAA acknowledges support by NSF grants AST-2009687 and AST-2108944, CXO grant TM2-23006X, Simons Foundation Award CCA-1018464, and Cottrell Scholar Award CS-CSA-2023-028 by the Research Corporation for Science Advancement.

\section*{Data Availability}

CAMELS data are publicly available at \url{https://camels.readthedocs.io/en/latest/}. Original data is available from the authors upon request by emailing \href{naomi.gluck@yale.edu}{naomi.gluck@yale.edu}.

\bibliographystyle{mnras}
\bibliography{main}

\appendix

\section{Additional Plots of Mock Datasets}
\label{sec:appendix_mock_datasets}

In this appendix, we provide additional maps and plots, including the scatter of $\Mhalo$ with total pixel counts per map in X-ray and HI (analogous to Fig.~\ref{fig:counts}), and Truth--Inference plots for inferred $\fcgm$, $\logZ$, $\fcool$, and $\logT$ for the HI+X-ray multifield with observational limits. We omit $\Mhalo$ or $\Mcgm$ here, as similar panels are shown in the diagonal panels of Fig.~\ref{fig:cross_sim_multi_Mhalo} and Fig.~\ref{fig:cross_sim_multi_Mcgm}. A summary of these properties and their trends with halo mass is shown in Fig.~\ref{fig:err_obslimit_type}.

\begin{figure*}
 
    \includegraphics[scale=0.44]{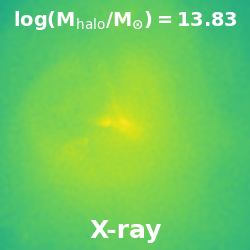}
    \includegraphics[scale=0.44]{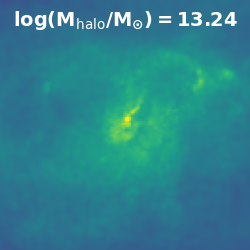}
    \includegraphics[scale=0.44]{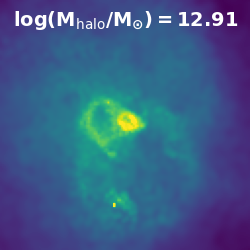}
    \includegraphics[scale=0.44]{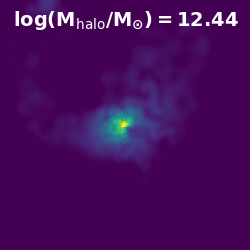}
    \includegraphics[scale=0.44]{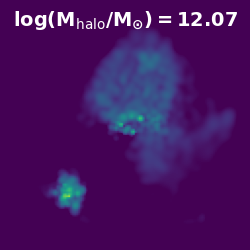}
    \includegraphics[scale=0.44]{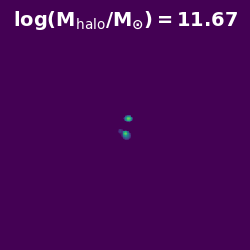}
    \\

    \includegraphics[scale=0.44]{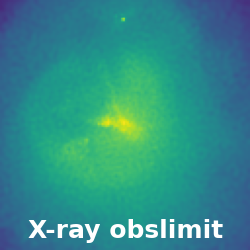}
    \includegraphics[scale=0.44]{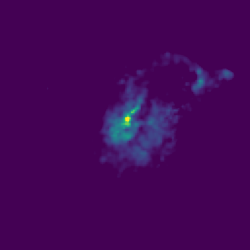}
    \includegraphics[scale=0.44]{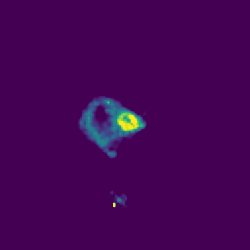}
    \includegraphics[scale=0.44]{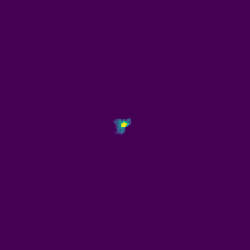}
    \includegraphics[scale=0.44]{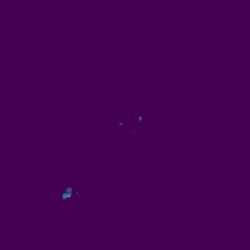}
    \includegraphics[scale=0.44]{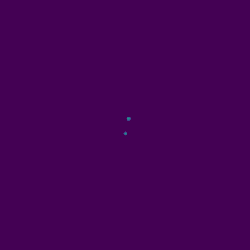}\\

    \includegraphics[scale=0.44]{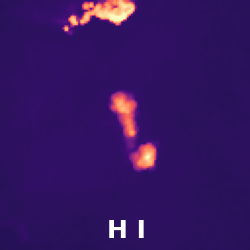}
    \includegraphics[scale=0.44]{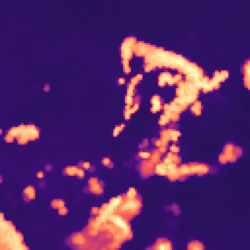}
    \includegraphics[scale=0.44]{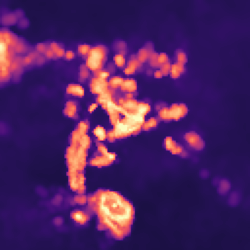}
    \includegraphics[scale=0.44]{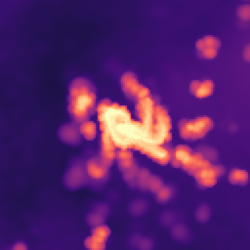}
    \includegraphics[scale=0.44]{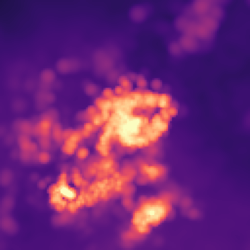}
    \includegraphics[scale=0.44]{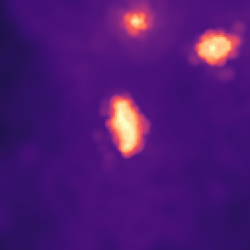}\\

    \includegraphics[scale=0.44]{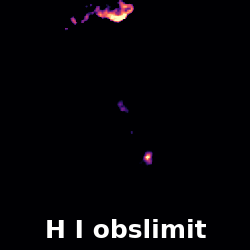}
    \includegraphics[scale=0.44]{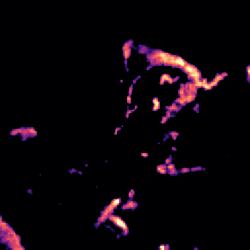}
    \includegraphics[scale=0.44]{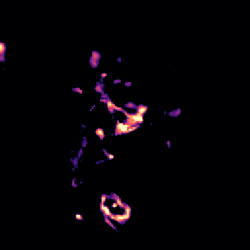}
    \includegraphics[scale=0.44]{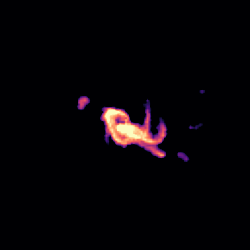}
    \includegraphics[scale=0.44]{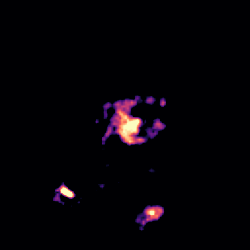}
    \includegraphics[scale=0.44]{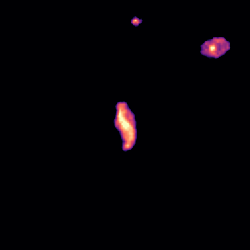}
    \caption{Maps of idealised X-ray (row 1), X-ray with observational limits (row 2), idealized HI (row 3), and HI with observational limits (row 4), as seen with IllustrisTNG. Moving across the row are halos of decreasing mass (approximately 0.5~dex), where columns correspond to the same halo map and hence the same mass.}
    \label{fig:maps_fn_mass_v2}
\end{figure*}

\begin{figure}
    \centering
    \includegraphics[width=0.48\textwidth]{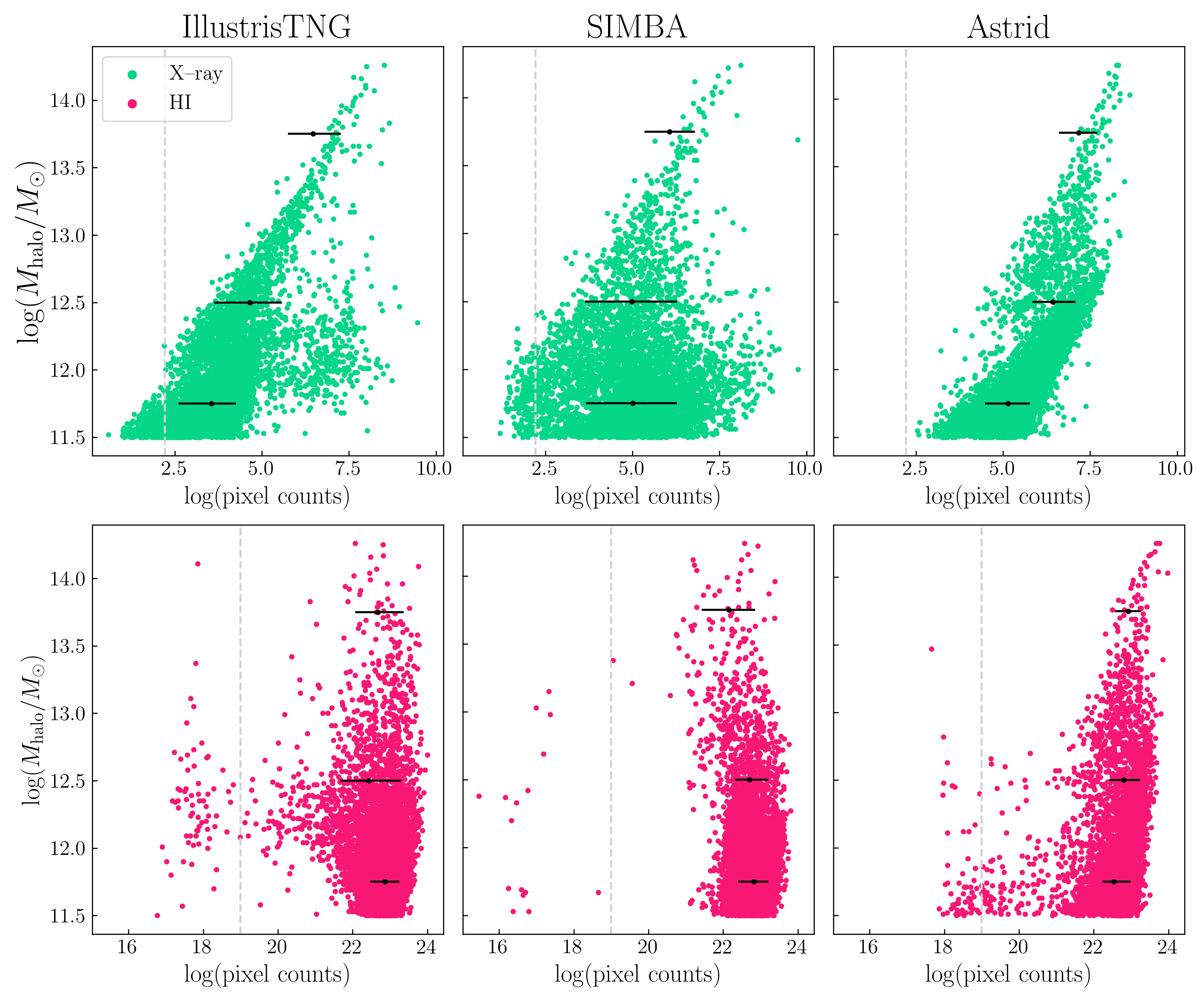}
    \caption{Halo mass as a function of the spatially integrated (total) flux in X-ray (top; green) and HI (bottom; pink) for \textit{all maps} available from the IllustrisTNG (left), SIMBA (middle), and Astrid (right) simulations. The vertical dashed line represents the observational limit of each field, and the black points represent the average value in each mass bin. The error bars represent the average 16-84 percentile in total flux for different halo mass bins. We see correlations only for IllustrisTNG and Astrid in X-ray.}
    \label{fig:mhalo_pix}
\end{figure}

Figure~\ref{fig:maps_fn_mass_v2} is an expanded version of Fig.~\ref{fig:all_sim_maps} for IllustrisTNG maps in X-ray (with and without observational limits, first and second rows, respectively) and HI (with and without observational limits, third and fourth rows, respectively) across most of the halo mass range explored in our analysis. Each column indicates four variations of the same halo. 

Figure~\ref{fig:mhalo_pix} illustrates the scatter of $\log(\Mhalo/M_{\odot})$, where each coloured point represents the total pixel value of each map along with the respective halo mass. ``Pixel counts”, as the total flux (X-ray) or the total column density (HI), are the sum of the pixels in each map (log-scaled). We only include one image axis (even though our CNN training set uses three rotations of the same halo along the three axes) so that the same halo does not appear more than once. The black points represent the average trends in each halo mass bin (see the definitions of the mass bin in Table~\ref{table:mass_bins}), and the error bars are the $16^{\rm th}$-$84^{\rm th}$ percentiles. Dashed grey vertical lines indicate the observational limits of each field, such that to the left of this line reside objects that would be too faint to observe with current instruments. The top row shows this scatter in X-ray, where there is a clear correlation with the halo mass for IllustrisTNG and Astrid. The bottom row shows the scatter in HI, similarly formatted. In this case, the correlations with halo mass for all simulations are either too weak or non-existent. Both fields match the expected trends from the visualisation of the maps in Fig.~\ref{fig:maps_fn_mass_v2}. This exercise aims to see if the halo mass can be predicted solely with total flux. Since the vertical scatter is not in the same order and is much larger than the network error, we cannot conclude that the halo mass is based only on the total flux.

\section{Additional Multi-field Truth-Inference Plots}
\label{sec:appendix_more_truth_inf_plots}

\begin{figure*}
    \centering
    \includegraphics[width=0.9\textwidth]{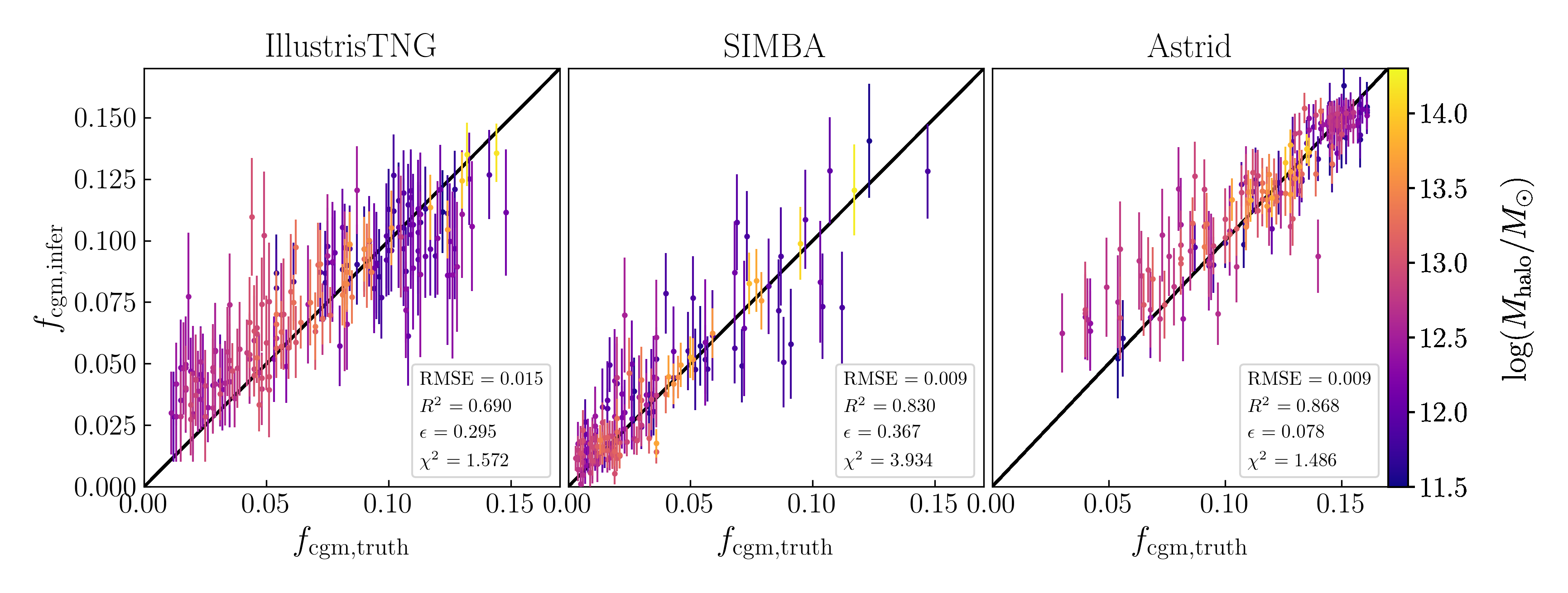}
    \caption{Truth--Inference plots for $\fcgm$ using HI+X-ray with observational limits for IllustrisTNG, SIMBA, and Astrid. These points are a fraction of the full dataset.}
    \label{fig:c1x3panel_fcgm}
\end{figure*}

\begin{figure*}
    \centering
    \includegraphics[width=0.9\textwidth]{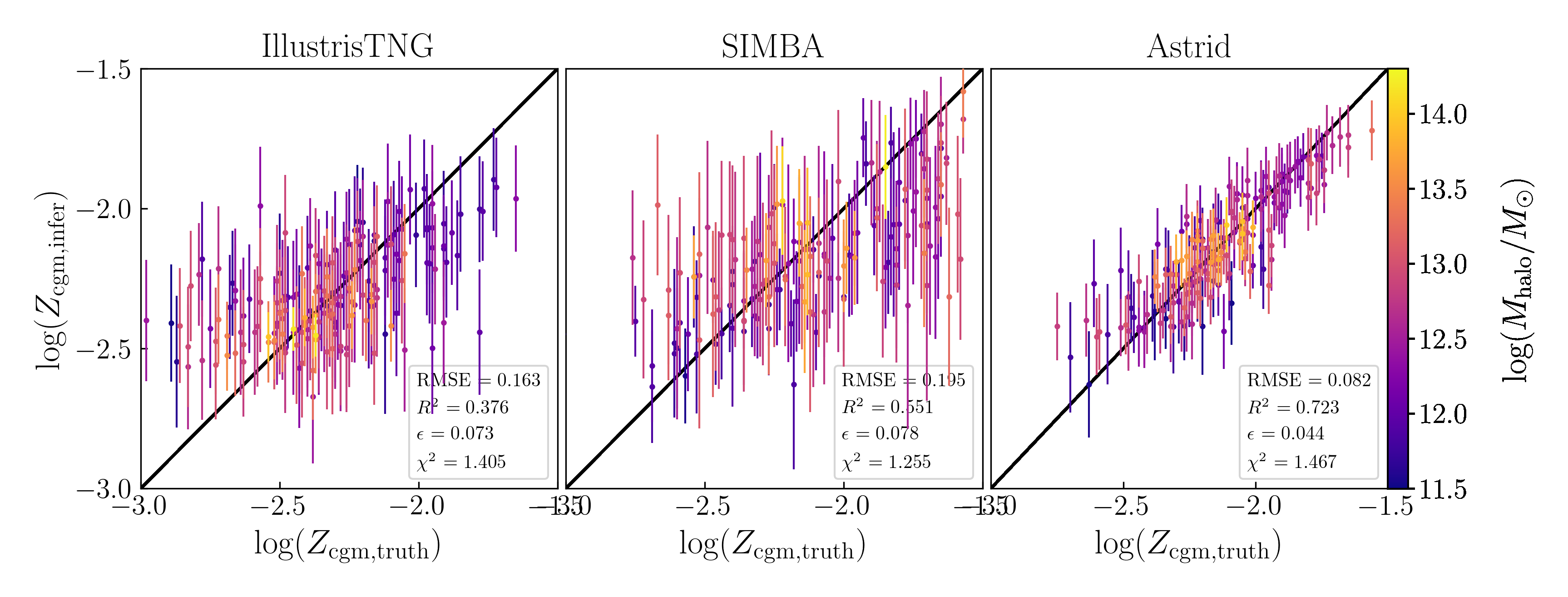}
    \caption{Truth--Inference plots for $\logZ$ using HI+X-ray with observational limits for IllustrisTNG, SIMBA, and Astrid. These points are a fraction of the full dataset.}
    \label{fig:c1x3panel_logZ}
\end{figure*}

\begin{figure*}
    \centering
    \includegraphics[width=0.9\textwidth]{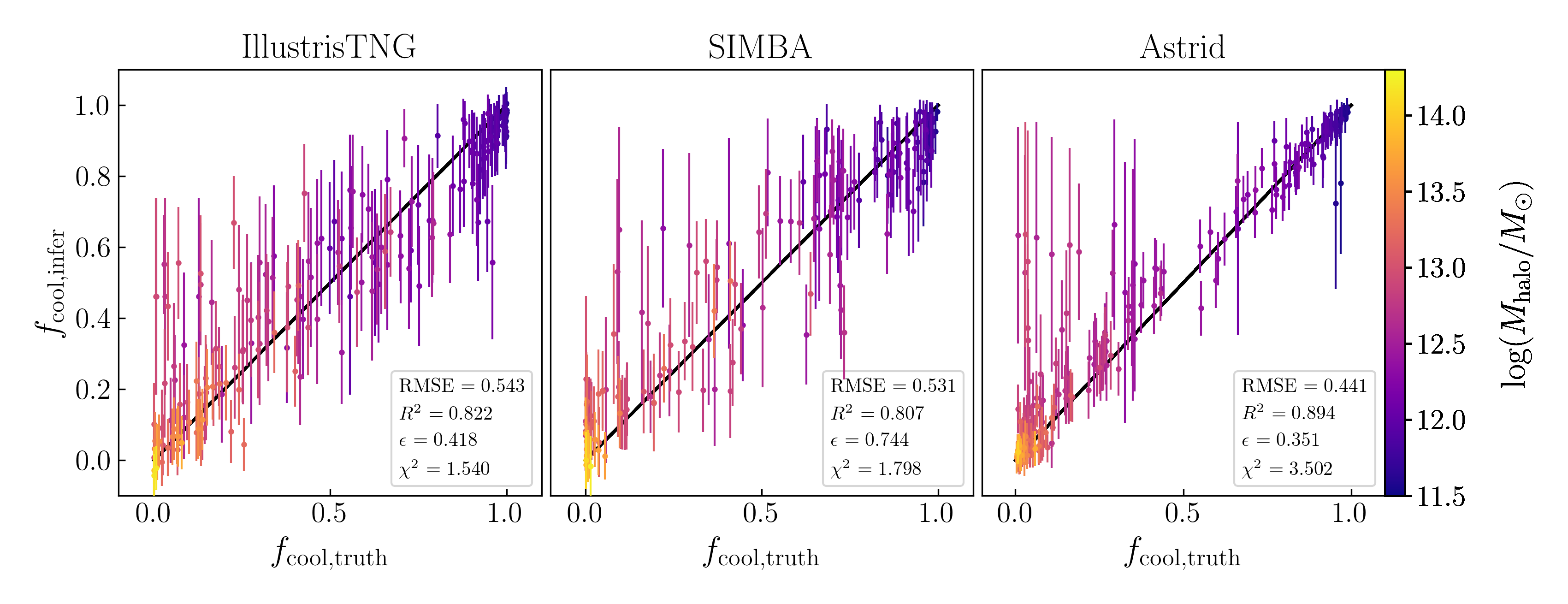}
    \caption{Truth--Inference plots for $\fcool$ (defined in Equation~\ref{eqn:fcool}) using HI+X-ray with observational limits for IllustrisTNG, SIMBA, and Astrid. These points are a fraction of the full dataset.}
    \label{fig:c1x3panel_fcool}
\end{figure*}

\begin{figure*}
    \centering
    \includegraphics[width=0.9\textwidth]{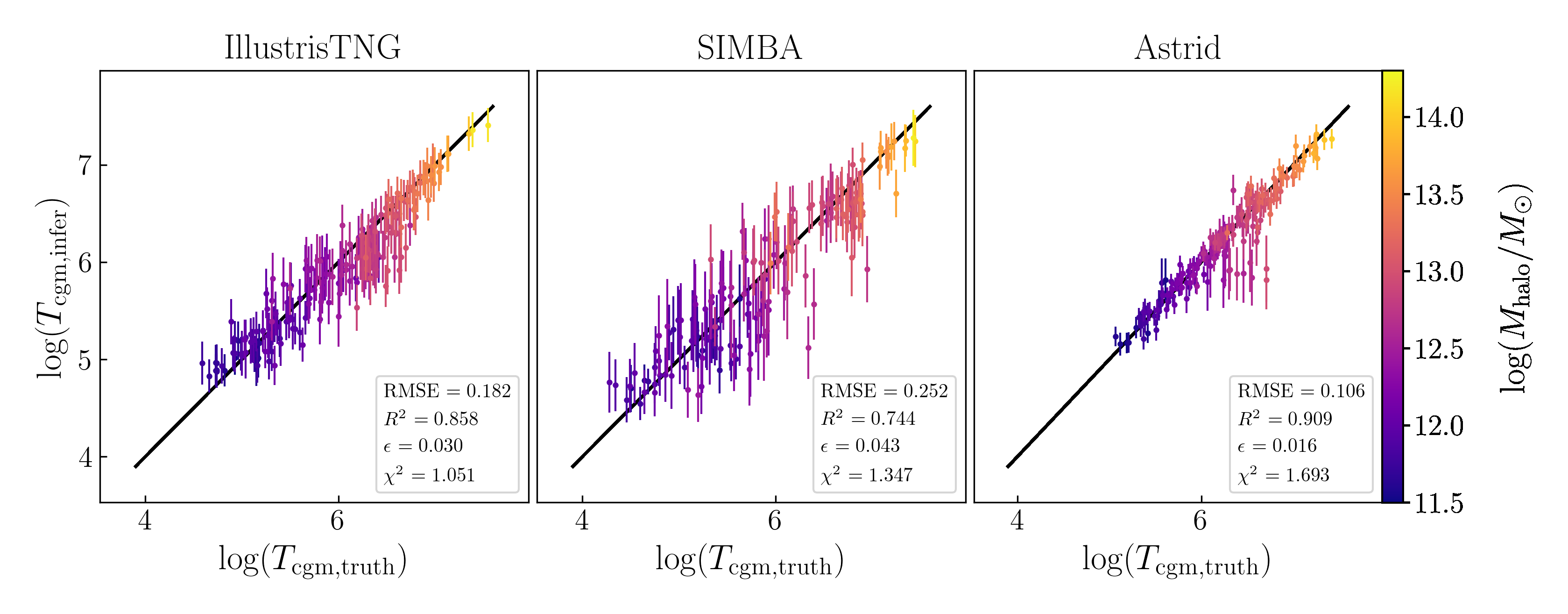}
    \caption{Truth--Inference plots for $\logT$ (defined in Equation~\ref{eqn:Tcgm}) using HI+X-ray with observational limits for IllustrisTNG, SIMBA, and Astrid. These points are a fraction of the full dataset.}
    \label{fig:c1x3panel_logT}
\end{figure*}

Figure~\ref{fig:c1x3panel_fcgm} shows the Truth--Inference plots for $\fcgm$ with the multi-field HI + X-ray and observational limits. We see a significant scatter throughout the mass range when training and testing on IllustrisTNG (left). Training and testing on SIMBA (middle) shows intermediate-mass halos clustered at low $\fcgm$ values, and low-mass halos scattered throughout. Training and testing on Astrid (right) shows a relatively low amount of scatter with small error bars, resulting in the lowest $\epsilon$ and highest $R^2$ value. Also, note that Astrid shifts the entire trend towards higher $\fcgm$ values and has its highest concentration of points towards higher values of $\fcgm$.

Figure~\ref{fig:c1x3panel_logZ} shows the Truth--Inference plots for $\logZ$ with the HI+X-ray multifield and observational limits. Training and testing on IllustrisTNG (left) or SIMBA (middle) results in significant scatter. Training and testing on Astrid shows a relatively low amount of scatter with small error bars. These trends look relatively similar to those with idealised maps of X-rays and HI (Fig.~\ref{fig:logZ_idealized_fields}). 

Figure~\ref{fig:c1x3panel_fcool} shows the Truth--Inference plots for $\fcool$ with the multi-field HI + X-ray and observational limits. All three simulations have inferences with significant scatter and large error bars. The multifield greatly improves the inference results for this property, which is difficult to constrain within the scope of this work. 

Finally, Figure~\ref{fig:c1x3panel_logT} shows the Truth--Inference plots for $\logT$ with the HI+X-ray multifield and observational limits. Training and testing on IllustrisTNG provide relatively good inference, with increased scatter for intermediate-mass halos. Training and testing on SIMBA show the largest scatter across the mass range. Training and testing on Astrid provides the least amount of scatter, with the smallest error bars and an overall impressive inference.

\section{CNN Architecture}
\label{sec:appendix_cnn_arch}

\begin{table}
\centering
\caption{Table outlining the main body of the convolutional neural network architecture used.}
\label{table:architecture}
\begin{tabular}{c c c c c c}
 \specialrule{.1em}{.05em}{.05em}
 \textbf{Layer} & \textbf{Input} & \textbf{Output} & \textbf{Kernel} & \textbf{Stride} & \textbf{Padding}\\ [0.25ex] 
 \specialrule{.1em}{.05em}{.05em}
 C01 & 1 & 12 & (3,3) & (1,1) & (1,1)  \\ 
 C02 & 12 & 12 & (3,3) & (1,1) & (1,1)  \\
 C03 & 12 & 12 & (2,2) & (2,2) & \\
 \hline
 \textbf{Layer} & \textbf{Size} & \textbf{$\epsilon$} & \textbf{Momentum} & \textbf{Affine} & \textbf{Tracking} \\ [0.25ex]
 \hline
 B01 & 12 & 1e-5 & 0.1 & True & True \\
 B02 & 12 & 1e-5 & 0.1 & True & True \\
 B03 & 12 & 1e-5 & 0.1 & True & True \\
 \hline
 C11 & 12 & 24 & (3,3) & (1,1) & (1,1) \\
 C12 & 24 & 24 & (3,3) & (1,1) & (1,1) \\
 C13 & 24 & 24 & (2,2) & (2,2) & -- \\
 B11 & 24 & 1e-5 & 0.1 & True & True \\
 B12 & 24 & 1e-5 & 0.1 & True & True \\
 B13 & 24 & 1e-5 & 0.1 & True & True \\
 C21 & 24 & 48 & (3,3) & (1,1) & (1,1) \\
 C22 & 48 & 48 & (3,3) & (1,1) & (1,1) \\
 C23 & 48 & 48 & (2,2) & (2,2) & -- \\
 B21 & 48 & 1e-5 & 0.1 & True & True \\
 B22 & 48 & 1e-5 & 0.1 & True & True \\
 B23 & 48 & 1e-5 & 0.1 & True & True \\
 C31 & 48 & 96 & (3,3) & (1,1) & (1,1) \\
 C32 & 96 & 96 & (3,3) & (1,1) & (1,1) \\
 C33 & 96 & 96 & (2,2) & (2,2) & -- \\
 B31 & 96 & 1e-5 & 0.1 & True & True \\
 B32 & 96 & 1e-5 & 0.1 & True & True \\
 B33 & 96 & 1e-5 & 0.1 & True & True \\
 C41 & 96 & 192 & (3,3) & (1,1) & (1,1) \\
 C42 & 192 & 192 & (3,3) & (1,1) & (1,1) \\
 C43 & 192 & 192 & (2,2) & (2,2) & -- \\
 B41 & 192 & 1e-5 & 0.1 & True & True \\ 
 B42 & 192 & 1e-5 & 0.1 & True & True \\ 
 B43 & 192 & 1e-5 & 0.1 & True & True \\ 
 C51 & 192 & 384 & (3,3) & (1,1) & (1,1) \\
 C52 & 384 & 394 & (3,3) & (1,1) & (1,1) \\
 C53 & 384 & 384 & (2,2) & (2,2) & -- \\
 B51 & 384 & 1e-5 & 0.1 & True & True \\ 
 B52 & 384 & 1e-5 & 0.1 & True & True \\ 
 B53 & 384 & 1e-5 & 0.1 & True & True \\ 
 C61 & 384 & 768 & (2,2) & (1,1) & -- \\
 B61 & 768 & 1e-5 & 0.1 & True & True \\
 \specialrule{.1em}{.05em}{.05em}
 
\end{tabular}
\end{table}

\begin{table}
    \centering
    \caption{A continuation of the neural network architecture, following Table~\ref{table:architecture}.}
    \label{table:arch_end}
    \begin{tabular}{c c c c c}
    \specialrule{.1em}{.05em}{.05em}
    \textbf{Layer} & \textbf{Type} & \textbf{Kernel Size} & \textbf{Stride} & \textbf{Padding} \\[0.25ex]
    \specialrule{.1em}{.05em}{.05em} 
    P0 & AvgPool2d & 2 & 2 & 0 \\
    \hline 
    \textbf{Layer} & \textbf{Type} & \textbf{Feat. In} & \textbf{Feat. Out} & \textbf{Bias} \\
    \hline
    FC1 & Linear & 768 & 384 & True \\
    FC2 & Linear & 384 & 12 & True \\
    \hline 
    \textbf{Layer} & \textbf{Function} & \textbf{p-value} & \textbf{in-place} & \textbf{Slope} \\
    \hline
    Dropout & \texttt{Dropout()} &0.3522 & False & \\

    ReLU & \texttt{ReLU()} & & & \\
    LeakyReLU & \texttt{LeakyReLU()} & & & -0.2 \\ 
    tanh & \texttt{Tanh()} & & & \\
    \specialrule{.1em}{.05em}{.05em}
    \end{tabular}
\end{table}

Initially, a similar CNN was applied to the CAMELS Multifield Dataset (CMD, \cite{camels_2021_w_ML}) as continuous 2D maps with the aim of constraining two cosmological parameters ($\sigma_8$ and $\Omega_M$), and four astrophysical feedback parameters ($A_{\rm{SN1}}$, $A_{\rm{SN2}}$, $A_{\rm{AGN1}}$, $A_{\rm{AGN2}}$) whose definitions change depending on the simulation used. Note that 3D maps are also available and can be reduced to obtain the existing 2D maps, but are not used for this analysis. A multifield allows the combination of fields to determine which singular or multiple fields return the tightest and most accurate constraints on any given parameter. The parameters currently available in the original network for the CMD are gas properties (density, velocity, temperature, pressure, metallicity), neutral hydrogen density, electron number density, magnetic fields, magnesium-ion fraction, dark matter density, and velocity, stellar mass density, and the total matter density.

We now define the variables used for the CNN used for this work in Table~\ref{table:architecture}. The names of layers beginning with C refer to \texttt{Conv2d}, and B refers to \texttt{BatchNorm2d}. Each type of layer has different input variables as described in the first mention of the layer type, with more details in the \cite{paszke_pytorch_2019} documentation for \texttt{PyTorch}. Subsequent layers of these two types do not have headings, but the numbers in the columns refer to the variable names and definitions when they are first mentioned. 

For the \texttt{Conv2d} layers, \texttt{Input} and \texttt{Output} are the size of the image produced as it passes through each layer. \texttt{Kernel} refers to the size of the kernel or the grid space in any particular layer. \texttt{Stride} is the number of rows and columns that have passed through each ``slide'' or translation between layers. If computational efficiency is not an issue, in some cases, it can be more accurate to slide one element at a time. However, cutting out the intermediate steps and increasing the stride for larger datasets like the one used here is more efficient. \texttt{Padding} refers to filling the kernel's edges after each layer. As the dimensions of the image decrease and eventually reach $1\times 1$, we need to fill the space left after each dimensional reduction. One padding mode is the ``zeros'', where values of 0 are used as a filler as the image is processed through the network. Another common mode is ``circular'', where the grid is filled with the value at the boundary of the image in the current stage.

For the \texttt{BatchNorm2d} layers, where \texttt{Size} refers to the number of features based on some expected input size from the previous layer. $\epsilon$ is added to the denominator of any value to ensure the stability of the pipeline and the results. \texttt{Momentum} can be set to None if a cumulative moving average (simple average) is being computed, but the default value is 0.1 for the running mean and running variance computations. Note that this argument is slightly different from what is generally used in optimiser classes. \texttt{Affine}, if set to True, allows weights and biases to be defined, which are $\gamma$ and $\beta$, respectively, within the documentation. \texttt{Tracking}, if set to True as the default, tracks the mean and variance. If set to False, statistics buffers are initialised such that \texttt{running-mean} and \texttt{running-var} is set to None, and the module uses only batch statistics for training and testing modes.

In Table~\ref{table:arch_end}, the functions mentioned directly follow those of Table~\ref{table:architecture}. The \texttt{P0: AvgPool2d} layer includes the \texttt{kernel size} for the window size. \texttt{Stride} and \texttt{Padding} have similar definitions as before, where \texttt{Stride} now refers to the stride of the window, where it defaults to the same value as the kernel size, and \texttt{Padding} is defaulted to ``zeros'' mode, discussed previously. 

A linear transformation is applied for both Fully Connected (FC) layers, where the \texttt{Feat. In} and \texttt{Feat. Out} refers to the size of each input and output sample, respectively. Setting the \texttt{bias} to True (as the default) allows the activation function to be shifted by some constant amount, known as the bias, to the layer input. The dropout layers randomly disengage some neurons with some probability, \texttt{p-value}, or just as \texttt{p}, to discourage some neurons from being favoured over others. The default p-value is 0.5. Finally, if set to True, \texttt{inplace} will randomly set the neurones to zero in place. The default value for this parameter is False, where the results of the dropout layer are saved to a separate variable to be potentially used later.

The \texttt{ReLU()}, or the rectified linear activation function (linear, piece-wise), takes this form:
\begin{equation}
    \rm{ReLU(x)} = (x)^+ = \rm{max}(0,x)
\end{equation}
The input is returned directly if positive and will be set to zero otherwise. The \texttt{LeakyReLU()} (Leaky rectified linear unit) is defined as

\begin{gather}\rm{LeakyReLU}(x) = \left\{
    \begin{aligned}
        &x, && \rm{if} x \ge 0 \\ & \rm{negative\_slope}\times x, && \rm{otherwise}
    \end{aligned}\right.
\end{gather}
where the \texttt{negative\_slope} controls the slope angle specifically used for negative input values. The default value is 0.001, such that instead of a flat slope for negative values, it has a small slope, determined before training begins, and is not a result of the training process. The \texttt{Tanh()} function is defined as,

\begin{equation}
    \rm{Tanh}(x) = \frac{\exp(x)-\exp(-x)}{\exp(x)+\exp(-x)}
\end{equation}
where it is used in place of the sigmoid function, as it is more computationally efficient for networks with multiple layers.


\bsp	
\label{lastpage}
\end{document}